\documentclass{article}
\usepackage[a4paper]{geometry}
\usepackage{amsmath}
\usepackage{amsthm}
\usepackage{tikz}
\usepackage{cases}
\usepackage{caption}
\usepackage{subcaption}
\usepackage{comment}
\usepackage{amssymb,latexsym}     
\usepackage{url}
\usepackage{xcolor}
\usepackage[ruled,vlined]{algorithm2e}
\usepackage{wrapfig}
\usepackage{stmaryrd}
\usepackage{graphicx}
\usepackage{float}
\graphicspath{{./img/}}
\theoremstyle{plain}%

\theoremstyle{definition}%

\makeatletter
\g@addto@macro\@floatboxreset\centering
\makeatother

\makeatletter
\renewcommand*{\@fnsymbol}[1]{\ifcase#1\or*\else\@arabic{\numexpr#1-1\relax}\fi}\makeatother

\definecolor{violet}{RGB}{73,0,146}      
\definecolor{ocean}{RGB}{0,109,219}      
\definecolor{bordeaux}{RGB}{146,0,0}     
\definecolor{marron}{RGB}{146,73,0}      
\definecolor{clementine}{RGB}{219,109,0} 
\definecolor{acier}{RGB}{0,73,73}        %
\definecolor{pin}{RGB}{0,146,146}        %
\newcommand{\rondrouge}{{\large{\color{clementine}$\bullet$}}}
\newcommand{\carrebleu}{{\small{\color{ocean}$\blacksquare$}}}
\newcommand{\trianglevert}{{\color{pin}$\blacktriangle$}}

\newcommand*{\DontPrintSemicolon}[0]{\dontprintsemicolon}

\title{
  \Large{Targeting  realistic geometry in Tokamak code \gysela}%
\footnote{This work has been carried out within the framework
    of the EUROfusion Consortium and has received funding from the Euratom
    research and training program 2014-2018 under grant agreement No
    633053. The views and opinions expressed herein do not necessarily reflect
    those of the European Commission.}}
\author{ Nicolas Bouzat \thanks{Inria, F-54600 Villers-l\`es-Nancy, France,
    hosted by CEA/IRFM, F-13108 Saint-Paul-lez-Durance, France}
  \and 
  Camilla Bressan \thanks{D-85748
    Garching, \& Technische Universit\"at M\"unchen, D-85748
    Garching, Germany.}
    \and 
  Virginie Grandgirard \thanks{CEA/IRFM, F-13108 Saint-Paul-lez-Durance,  France}
  \and 
  Guillaume Latu $^{3}$
  \and 
  Michel Mehrenberger \thanks{IRMA, Universit\'e de Strasbourg, FR-67084
    Strasbourg \& Inria, F-54600 Villers-l\`{e}s-Nancy, France}
  }

\def\NN{\mathbb{N}}

\newcommand{\vecto}[1]{\overrightarrow{#1}}
 \let\pa\partial

\newcommand{\xg}{\mathcal{X}_G}
\newcommand{\vpar}{v_{\parallel}}

\newcommand{\gysela}{\textsc{Gysela}}

\newcommand{\Bstar}{B_{\|}^{\ast}}

\newcommand{\fbar}{\bar{f}}

\newcommand{\dv}{\,{\text{d}}{\mathbf{v}}}

\newcommand{\Ntheta}{N_{\theta}}
\newcommand{\Dtheta}{\Delta \theta }
\newcommand{\Dthetai}[1]{\Delta \theta_{[#1]}}
\newcommand{\Nthetai}[1]{N_{\theta_{[#1]}}}
\newcommand{\ff}{f}
\newcommand{\fftilde}{\tilde{f}}
\newcommand{\ffij}{f_{i,j}}
\newcommand{\ffi}[1]{f_{i,{#1}}}

\newcommand{\rmin}{r_{\rm min}}
\newcommand{\rmax}{r_{\rm max}}
\newcommand{\Rij}{R_{i,j}}
\newcommand{\TLambda}{\tilde{\Lambda}}

\begin{document}
\maketitle

\begin{abstract} 
In magnetically confined plasmas used in Tokamak, turbulence is
responsible for specific transport that limits the performance of this
kind of reactors. Gyrokinetic simulations are able to capture ion and
electron turbulence that give rise to heat losses, but require also
state-of-the-art HPC techniques to handle computation costs.  Such
simulations are a major tool to establish good operating regime in
Tokamak such as ITER, which is currently being built.  Some of the
key issues to address more realistic gyrokinetic simulations are:
efficient and robust numerical schemes, accurate geometric
description, good parallelization algorithms. The framework of this
work is the Semi-Lagrangian setting for solving the gyrokinetic Vlasov
equation and the \gysela\  code. In this paper, a new variant for the
interpolation method is proposed that can handle the mesh singularity
in the poloidal plane at $r=0$ (polar system is used for the moment in
\gysela). A non-uniform meshing of the poloidal plane is proposed instead of
uniform one in order to save memory and computations. The interpolation
method, the gyroaverage operator, and the Poisson solver are revised
in order to cope with non-uniform meshes. A mapping that establish a
bijection from polar coordinates to more realistic plasma shape is used to
improve realism. Convergence studies are provided to establish the
validity and robustness of our new approach.
\end{abstract}


\section*{Introduction}
\label{sec:intro}
Understanding and control of turbulent transport in thermonuclear plasmas in
magnetic confinement devices is a major goal. This aspect of first principle
physics plays a key role in achieving the level of performance expected in
fusion reactors. In the ITER design\footnote[6]{http://www.itercad.org/}, the
latter was estimated by extrapolating an empirical law. The simulation and
understanding of the turbulent transport in Fusion plasmas remains therefore an
ambitious endeavor.

The Fusion energy community has been engaged in high-performance
computing (HPC) for a long time.  For example, gyrokinetic simulations
are time-hungry (thousands up to millions of CPU-hours) and we then need
large amount of computational time that are typically provided by
advanced computational facilities.  Computer simulation is and will
continue to be a key tool for investigating several aspects of Fusion
energy technology, because right now there is no burning plasma
experiments like ITER. Some of the key issues to address realistic
simulations of the Tokamak are: efficient and robust numerical
schemes, accurate geometric description, good parallelization
algorithms.

The \textit{gyrokinetic} framework considers a computational domain in
five dimensions (3D in space describing a torus geometry, 2D in
velocity). Time evolution of the system consists in solving Vlasov
equation non-linearly coupled to a Poisson equation (electrostatic
approximation, quasi-neutrality is assumed). The code has the
originality to be based on a semi-Lagrangian scheme\,\cite{sonnen99} and
it is parallelized using an hybrid OpenMP/MPI
paradigm\,\cite{crous09,latu07}.

Let $\vec{z}=(r,\theta,\varphi,\vpar,\mu)$ be a variable describing
the 5D phase space. The time evolution of the ionic distribution
function of the guiding-center $\fbar(\vec{z})$ (main unknown) is
governed by the gyrokinetic Vlasov equation (simplified version without right-hand side terms):
\begin{equation}
  \partial_t\fbar + \frac{1}{\Bstar}\nabla_{\vec{z}}\cdot\left(\frac{{\rm d}\vec{z}}{{\rm d}t}\Bstar\fbar\right) = 0
\label{vlasov_intro}
\end{equation}

The guiding-center motion described by the
previous Vlasov/transport equation is coupled to a field solver (3D quasi neutral solver
which is a Poisson-like solver) that computes the electric potential
{\footnotesize$\phi(r,\theta,\varphi)$} (adiabatic electron limit):
\begin{equation}
\frac{e}{T_e}(\phi-{\langle\phi\rangle)} = \frac{1}{n_0}\int{}J_0(\bar{f} - \bar{f}_{init})\dv + \rho_i^2\nabla_\perp^2 \frac{e\phi}{T_i}
\label{poisson_intro}
\end{equation}\\[-3mm]

We will not describe this last equation (details can be found in
\cite{virginie,latu11b}). This Poisson-like equation gives the
electric field $\phi$ that corresponds to the particle distribution at
each time step~$t$. The derivates of $J_0\,\phi$ along the torus dimensions
are computed. Then, these quantities act as a feedback in
the Vlasov equation, they appear into the term $\frac{{\rm d}\vec{z}}{{\rm
    d}t}\Bstar\fbar$. The Vlasov solver 
represents the critical CPU part, \textit{i.e.}  usually more than 90\%
of computation time.
This equation is solved by splitting it into the
advection equations ($\mathcal{X}_G=(r,\theta)$):

\vspace*{-0.1cm}
\noindent\begin{eqnarray*}
&\Bstar \pa_t\fbar+{\vecto{\nabla}}\cdot{\left(\Bstar{}\frac{d\xg}{dt}\fbar\right)}=0\ \ (\hat{\mathcal{X}_G}\ \ \textrm{operator}),\ \ \\
&\Bstar \pa_t\fbar+\pa_\varphi\left(\Bstar{}\frac{d\varphi}{dt}\fbar\right)=0\ \ (\hat{\varphi}\ \  \textrm{operator}),\ \ \ \
\Bstar \pa_t\fbar+\pa_{\vpar}\left(\Bstar{}\frac{d\vpar}{dt}\fbar\right)=0 \ \ (\hat{v_{\parallel}}\ \  \textrm{operator}).
\end{eqnarray*}
Each advection consists in applying a shift operator along one or two dimensions. A Strang splitting procedure is
employed to reach second order accuracy in time.
The sequence we choose is
$(\hat{v_{\parallel}}/2,\hat{\varphi}/2,\hat{\mathcal{X}_G},\hat{\varphi}/2,\hat{v_{\parallel}}/2)$,
where the factor $1/2$ is a shift over a reduced time step
$dt/2$.

In this work, we will propose solutions to improve the \scalebox{0.9}{$\hat{\mathcal{X}_G}$}
operator in the Vlasov solver, the gyroaverage $J_0$ that appears in
Eq. (\ref{poisson_intro}), and the 2D Poisson equation we need to solve that comes from
$\rho_i^2\nabla_\perp^2 \frac{e\phi}{T_i}$ term in
Eq. (\ref{poisson_intro}).
These three operators are tightly coupled to the geometry in the poloidal plane which is perpendicular (transverse) to the magnetic field direction.
Conversely, the \scalebox{0.9}{$\textstyle\hat{v_{\parallel}}$} and \scalebox{0.9}{$\textstyle\hat{\varphi}$} operators are quite independent from the poloidal geometry because they act in other dimensions than \scalebox{0.9}{$\textstyle \hat{\mathcal{X}_G}$}.\\
The paper is organized as follows: in the first section, the original poloidal
geometry and meshing is described, the new non-uniform approach focusing the poloidal plane is explained, and the mapping that handles realistic geometry is given.  Then, in the second section, 
the interpolation method on non-uniform polar mesh is investigated,
but also advection and gyroaverage operators on such a mesh. Also,
we show the numerical method chosen for the 2D Poisson solver. Finally, numerical results and convergence studies are presented in the third section.

\section{New geometry and mapping}
\label{sec:geometry}
Changing the mesh of the poloidal plane while keeping a polar coordinate system
should allow us first, to loosen the meshing in order to reduce the typical
concentration of points near the center $r=0$ and second, to have the mesh match
more closely the magnetic surfaces of the plasma. We should then have an
improvement in execution time by reducing the overall number of points as well
as an improvement in accuracy thanks to the grid being closer to the typical
pattern of simulated phenomena. The non-uniform meshing will also allows us to
focus on a specific location of the plane that we want to solve by using more
points there and only solving roughly elsewhere.

\subsection{Polar mesh}

\subsubsection{Original polar mesh}\label{sec:uniform}
We fix $N_r$, the number of points in the radial direction and $N_\theta$, the
number of points in the poloidal direction. The original polar mesh, as it is
defined in \gysela, is such as $r_i=r_{min}+i\Delta{}r$ with $r_{min}>0$,
$i\in\llbracket 0,N_r-1\rrbracket$, \mbox{$\Delta r =
  \frac{r_{\max}-r_{\min}}{N_r-1}$}, and also
$\theta_j=\frac{j\,2\pi}{N_{\theta}}$ with
$j\in\llbracket 0,N_\theta-1\rrbracket$. It is worth noting that $r_{min}$ and
$r_{max}$ act as boundary conditions. For each operator that is applied within
the poloidal domain, specific ad-hoc approaches are setup to handle what is
happening in the central hole $r\in[0,r_{min}]$. We will not detail the set of
ad-hoc boundary conditions that are described in \cite{grandgirard16}.

\subsubsection{New non-uniform polar mesh}
\label{sec:non_uniform}
The new poloidal grid that we want to use is sketched in
Figure~\ref{fig:newgrid}.  The idea is to have, for each different circle
labeled by $r$ coordinate, a different number of point in the radial direction
$\theta$. For instance, in Figure \ref{fig:newgrid}~(p.~\pageref{fig:newgrid}), the first layer (inner
circle) has four points, the second to fourth layers have eight points and the
remaining layers have sixteen points. This allows either to have a density of
grid point which is nearly uniform on the plane, or to model finely a subset of
the plane which is better solved with more grid points. 
This meshing or quite similar approaches have already been used in a set of 
papers\,\cite{Mohseni2000787,Holman2015ASF,Mock2011}. 
However, in these previous works the setting and the equations solved were quite 
different from what we investigate here. Therefore, we have mainly only retained the meshing
strategy while redesigning the operators and tools that apply on the mesh.

We have $r_{\min} = \frac{\Delta r}{2}$ and
$\ r_{\max}\!=\!r_{\min}+(N_r-1)\Delta r$ so that $\Delta
r\!=\!\frac{r_{\max}-r_{\min}}{N_r-1}=\frac{r_{\max}-\frac{\Delta
    r}{2}}{N_r-1}$, which leads to
$$
\Delta r\!=\!\frac{r_{\max}}{N_r-\frac{1}{2}},
$$
and the radial points are
$$
r_i\!=\!r_{\min}+i\,\Delta r\!=\!(i+\frac{1}{2})\Delta r,\quad i=0,\dots,N_r-1.
$$


Now, for each one of the $r_i$ we choose a number of points along $\theta$: $N_{\theta_{[i]}}$, and a grid spacing: $\Delta_{\theta\,[i]}$, according to what we
want to do. Either to focus on a specific region of the plane or to reduce the
overall number of points used on the plane and keep the same accuracy.

\subsection{Mapping}
\label{sec:mapping}
The previous approach can be combined with a general mapping, the polar mapping being only a special case. We focus here on mappings with analytical formula and whose inverse can also be expressed
by a formula (to shorten execution time) which was one of the concluding points of\,\cite{abiteboul2011solving}. This is of course the case for the polar mapping, but we can also find other more general cases, that can have relevance for the description of the geometry of a tokamak.
 We consider here the case of a large aspect ratio Tokamak equilibrium, and the mapping that derives from it, as in~\cite{fitzpatrick92,angelino08}. 


\noindent For the polar mapping
$$
x=r\cos(\theta),\ y = r\sin(\theta),
$$
the  inverse mapping is given by
$$
r = \sqrt{x^2+y^2},\ \theta= {\rm atan2}(y,x).
$$


\noindent For the large aspect ratio mapping (see  \cite{angelino08}; the formula is similar, only $\omega$ is changed into $\pi-\omega$), we have the formula
$$
\begin{array}{l}
x = R_0+r\cos(\omega)-\delta(r)-E(r)\cos(\omega)+T(r)\cos(2\omega)-P(r)\cos(\omega)\\ 
y = r\sin(\omega)+E(r)\sin(\omega)-T(r)\sin(2\omega)-P(r)\sin(\omega),
\end{array}
$$
where $\delta,\ E,\ T$ stand for Shafranov shift, elongation and triangularity. The $P$ notation corresponds to a relabeling of the surfaces. We refer to \cite{angelino08}, for the physical interest of such mapping in the tokamaks plasma community. We take here $P=0,\ T=0,\ \omega=\theta$, together with $E(r)=E_0 r$ and $\delta(r) = \delta_0r^2$; this clearly restrict the range of geometries, but enables to 
get an explicit formula for the inverse.
We get
\begin{equation}
\begin{array}{l}
x=R_0-\delta_0r^2+(1-E_0)r\cos(\theta)\\ 
y =(1+E_0) r\sin(\theta).
\end{array}
\label{equation_culham}
\end{equation}
The inverse mapping can be explicitly given.
Putting
$$
\tilde{y} = \frac{y}{1+E_0},\ \tilde{x} = \frac{x-R_0}{1-E_0},\ \tilde{\delta}_0 = \frac{2\delta_0}{1-E_0},
$$
we are lead to solve
$$
\left(x+\frac{\tilde{\delta}_0}{2}r^2\right)^2+y^2=r^2.
$$
We find

$$
r = \left(\frac{2\left(\tilde{x}^2+\tilde{y}^2\right)}
{1-\tilde{\delta}_0\tilde{x}
+\sqrt{(1-\tilde{\delta}_0\tilde{x})^2-\tilde{\delta}_0^2(\tilde{x}^2+\tilde{y}^2)}}\right)^{1/2},
\ \theta =  {\rm atan2}(\tilde{y},\tilde{x}+\frac{\tilde{\delta}_0}{2}r^2).
$$

\noindent Note that $r$ is well defined as soon as
$
\tilde{\delta}_0\left(\tilde{x}+\sqrt{\tilde{x}^2+\tilde{y}^2}\right)\le 1,
$
and when $\delta_0=E_0=R_0=0$, we recover the polar mapping\footnote{This is not the case for the other solution of the polynomial of degree $2$ in $r^2$:
$r=\left(\frac{2\left(\tilde{x}^2+\tilde{y}^2\right)}{1-\tilde{\delta}_0\tilde{x}-\sqrt{(1-\tilde{\delta}_0\tilde{x})^2-\tilde{\delta}_0^2(\tilde{x}^2+\tilde{y}^2)}}\right)^{1/2}$
}.
We refer to \cite{abiteboul2011solving} (there, the inverse mapping is also needed) and \cite{hamiaz2015semi,HamiazBack} for some works concerning 
the semi-Lagrangian method combined with a mapping. 
In the following, we will take  
\begin{equation}
E_0=0.3, R_0=0.08,\ \delta_0=0.2.
\label{param_culham}
\end{equation}
Figure \ref{fig:mapping_large_ar} (page \pageref{fig:mapping_large_ar}) shows a non-uniform grid combined with this specific mapping.


\section{Operators in complex geometry}
\label{sec:operators}
\subsection{Lagrange interpolation in 2D}
\label{sec:interp}
Let first consider a uniform mesh to introduce the notations, \textit{i.e}
$\forall i\in\llbracket 0,N_r-1\rrbracket, \Nthetai{i}=\Ntheta{}$. Let suppose that
$r\in [\frac{\Delta r}{2}, r_{\text{max}}[$, $\theta\in [0,2\pi[$, but also let
        us define $h$ and $k$ in $\llbracket 0,N_r-1 \rrbracket$ and $\llbracket
        0,\Ntheta -1 \rrbracket$ such as $r_h \le r < r_{h+1}$ and $\theta_k \le
        \theta < \theta_{k+1}$ where $r_h = (h+\frac{1}{2})\Delta r$ and
        $\theta_k = k\Dtheta{}$. Given an order of interpolation $p-1$, the
        Lagrange interpolation polynomial equals
\begin{equation}
  \label{eq:lagrange_poly}
  \quad L^{(p)}(r,\theta) = \sum_{m=l}^{u}\sum_{n=l}^{u}f(r_{h+m},\theta_{k+n})L^{(p)}_{m,n}(r,\theta)
\end{equation}
where $l=-\lfloor\frac{p-1}{2}\rfloor$, $u=\lfloor\frac{p}{2}\rfloor$ and
$L^{(p)}_{m,n}$ are the Lagrange basis polynomial. 
Then the basis polynomial $L^{(p)}_{m,n}(r,\theta)$ associated to the
point ($r_{h+m},\theta_{k+n}$) reads
\begin{equation}
  \label{eq:coeff_lagrange}
  L^{(p)}_{m,n}(r,\theta) = 
  \prod_{\substack{i=l \\ i\ne m}}^{u} (
  \frac{r-r_{h+i}}{r_{h+m}-r_{h+i}} )\times
  \prod_{\substack{j=l \\ j\ne n}}^{u} (
  \frac{\theta-\theta_{k+j}}{\theta_{k+n}-\theta_{k+j}}).
\end{equation}

\noindent We can defined a unique set of $(\beta,\gamma)$ such as $r = r_h + \beta \Delta r$ and
$\theta = \theta_k + \gamma \Dtheta{}$, with $(\beta,\gamma)\in[0,1[^2$. Then
Eq. (\ref{eq:coeff_lagrange}) can be simplified to
\begin{equation*}
  L^{(p)}_{m,n}(r,\theta) = \prod_{\substack{i=l \\ i\ne m}}^{u} (
  \frac{\beta - i}{m-i}) \times \prod_{\substack{j=l \\ j\ne n}}^{u} (
  \frac{\gamma-j}{n-j}). 
\end{equation*}

If the radial position $r$ goes above $r_{max}$ then the coefficient
is computed the same way but a Dirichlet condition is used and
$f(r,\theta)$ is cast to $f(r_{max},\theta)$. If the radial position
is located in the interval~$r\in[0,\frac{\Delta{}r}{2}[$, the interpolation scheme has to be
adapted because we are crossing the most inner radius of the grid. Let
suppose, we have $h+m<0$ in Eq. (\ref{eq:lagrange_poly}), we have to recast $r$ and
$\theta$ coordinates at the same time to cross the center at
$r=0$. The new coordinates of a mesh point with $h+m<0$ located at $(r_{h+m},\theta_{s})$  are set to
$(r_{-h-m-1},\theta_{s+\frac{\Ntheta{}}{2} [\Ntheta{}]})$. We basically continue
the stencil on the radially opposite side of the grid by performing a $\pi$ rotation.


\begin{figure}
  \begin{minipage}[b]{0.3\textwidth}
    \centering
    \includegraphics[trim = 0cm 0cm 0cm 0cm,clip,height=3.6cm]{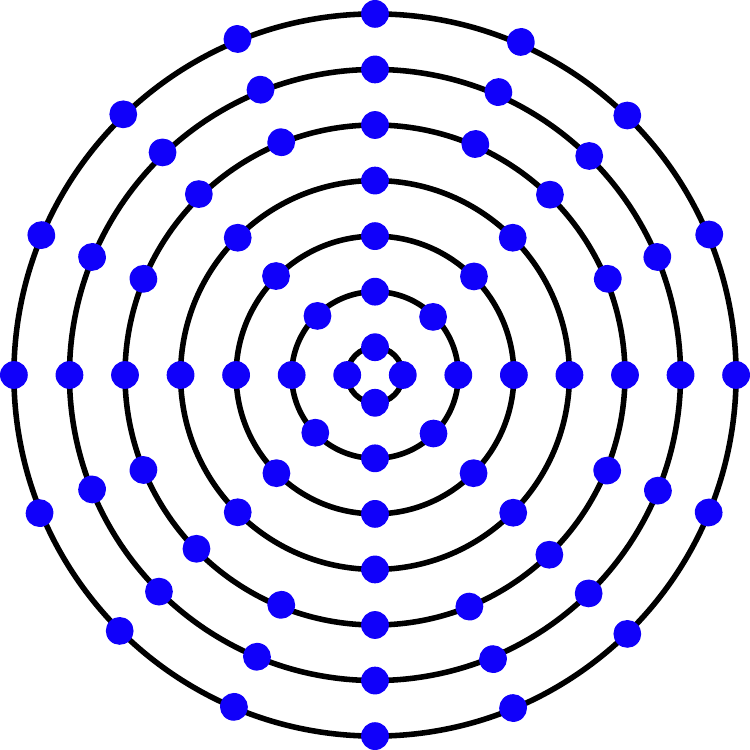}
    \vspace*{1.03cm}
    \caption{New poloidal grid. The number of points in $\theta$ direction
      (angle) depends on the radial position (distance to the center).}
    \label{fig:newgrid}
  \end{minipage}\hspace*{3mm}%
  \begin{minipage}[b]{0.3\textwidth}
    \begin{center}
      \includegraphics[trim = 0cm 0cm 0cm 0cm,clip,height=4.3cm]{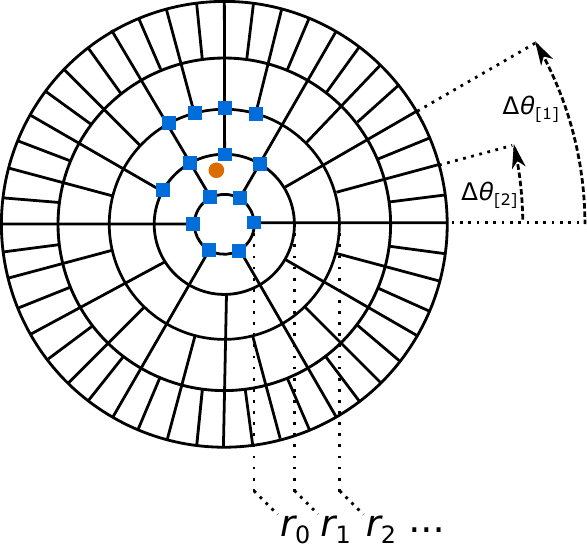}
    \end{center}
    \caption{Interpolation of a point~\rondrouge~of the poloidal plane with a
      stencil of 16 points~\carrebleu~(Lagrange of \mbox{order~3}).}
    \label{fig:interp}
  \end{minipage}\hspace*{3mm}%
  \begin{minipage}[b]{0.3\textwidth}
    \begin{center}
      \includegraphics[trim = 0cm 0.1cm 0cm .2cm,clip,height=5cm]{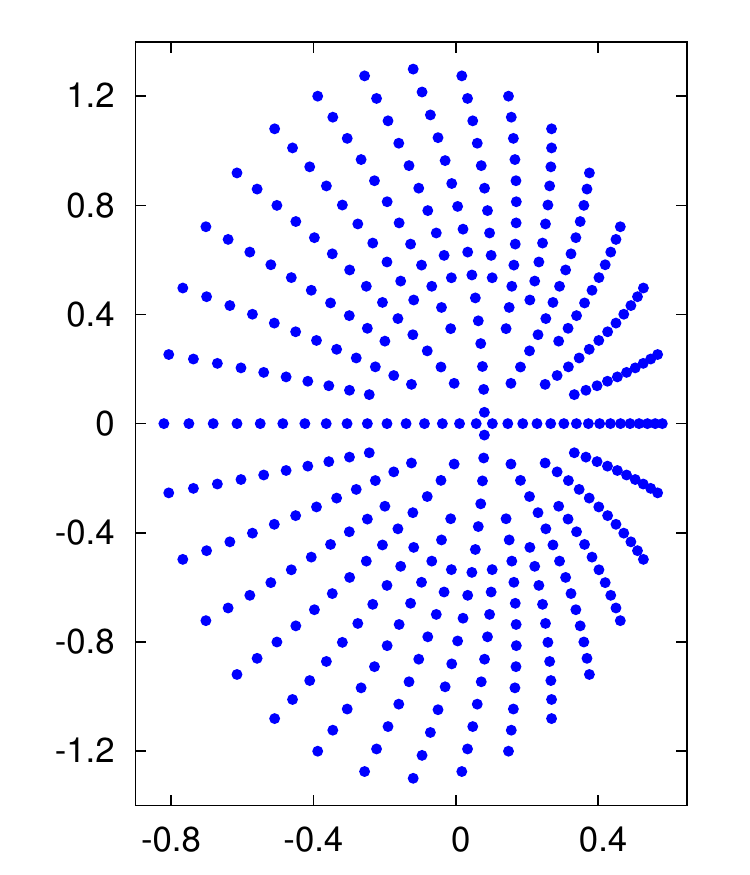}
    \end{center}
    \caption{New mapped grid using large aspect ratio equilibrium, with non-uniform meshing along $\theta$.\\
\ }
    \label{fig:mapping_large_ar}
  \end{minipage}
\end{figure}

Considering a non-uniform mesh as described in
Section~\ref{sec:geometry}, we need to take into account the cases where the
interpolation stencil covers several radii as shown in Fig.~\ref{fig:interp}. In
that case, the number of points along~$\theta$ for each radius may be different
and so the index of the nearest point in~$\theta$ direction may be
different. For instance on this Figure, the indexes of the interpolation points
on radius $r_1$ at ($1.5\,\Delta r$) are $2$, $3$, $4$, $5$ ($2\Dthetai{1}$,
$3\Dthetai{1}$, $4\Dthetai{1}$, \dots) and on radius $r_2$ ($2.5\,\Delta
r$) these indexes are $5$, $6$, $7$, $8$
($5\Dthetai{2}$,$6\Dthetai{2}$,\dots). In this way, we always use the closest
known points, leading to a good accuracy. To adapt the interpolation calculation
to non-uniform meshes, with notation of Eq. (\ref{eq:lagrange_poly}), one only
need to first perform the interpolation in $\theta$ before the one along~$r$. It allows to easily take into account that the number
$\Nthetai{i}$ depends on radius~$r_i$. 
Indeed, if the first interpolation was along
$r$, points along the
$\theta$ direction would possibly not be available (there is possibly not the same number of
points in
$\theta$ for each radius) and it would require extra 1D interpolations along
$\theta$ to fix this problem. Algorithm~\ref{alg:interp} summarizes how the 2D
Lagrangian interpolation is performed for non-uniform meshes.

\begin{algorithm}
  \DontPrintSemicolon
  \KwData{$f:\llbracket 0,N_r-1 \rrbracket \times \llbracket 0,\Nthetai{i}-1
    \rrbracket \rightarrow \mathbb{R}$, distribution function\\
    \qquad \quad~$(r,\theta)$, coordinates of the point \\
    \qquad \quad~$p$, degree of interpolation.}  
  \KwResult{$value$, interpolation of $f(r,\theta)$.}
  \Begin{
    $value = 0$ \\
    $h=\lfloor\frac{r-\Delta r/2}{\Delta r}\rfloor$;\quad $\beta = \frac{r - (\Delta r/2 +
      h\Delta r)}{\Delta r}$ \tcc*[f]{Radial position : $r=(h+\beta)\Delta r$}\\
    $l=-\lfloor\frac{p-1}{2}\rfloor$;\quad $u=\lfloor\frac{p}{2}\rfloor$ \\
    \For{$m\in\llbracket l,u \rrbracket$}{
      $c_r = 1$\\
      \For(\tcc*[f]{Computation of radial coefficient}){$i\in\llbracket l,u \rrbracket,\, i \ne m$}{
        $c_r = c_r \times \frac{\beta-i}{m-i}$
      }
      $h' = (h+m)$\\ 
      $k=\lfloor\frac{\theta}{\Dthetai{h'}}\rfloor$;\quad $\gamma =
      \frac{\theta - (k\Dthetai{h'})}{\Dthetai{h'}}$ \tcc*[f]{Poloidal position on radii $m$}\\
      \For{$n\in\llbracket l,u \rrbracket$}{
        $c_\theta = 1$\\
        \For(\tcc*[f]{Computation of poloidal coefficient}){$j\in\llbracket l,u \rrbracket,\, j \ne n$}{
          $c_\theta = c_\theta \times \frac{\gamma-j}{n-j}$
        }
        $k' = (k+n)$\\ 
        $(h'',k'') = $\texttt{ get\_plane\_indexes}$(h',k')$ \tcc*[f]{Radial boundary conditions}\\
        $value = value + f(h'',k'')\times c_r \times c_\theta$
      }
    }
  }
  \caption{Lagrange interpolation, tensor product  in 2D}
  \label{alg:interp}  
\end{algorithm}

\subsection{Gyroaverage operator} \label{sec:gyroaverage}
The gyroaverage operator is a key element in solving the Vlasov-Poisson system of equations, since it allows for the transformation of the guiding center distribution into the actual particle distribution, thus reducing the dimensionality of the system of one. The cyclotronic motion of the particles around the magnetic field lines at a distance below the Larmor radius is neglected without loss of accuracy, since this motion is much faster than the turbulence effects usually investigated; moreover, even modern computational power doesn't allow for such highly costly simulations.

A gyroaverage operator has been constructed on the new grid, and we will here briefly describe the numerical implementation which has been adopted in this context.

The gyroaverage operator depending on the spatial coordinates in the polar plane is defined as follows \cite{gyroaverage_steiner}:
\begin{equation}\label{eq:gyroaverage_definition}
\textit{J}_\rho(f)(r,\theta) = \frac{1}{2 \; \pi} \int_0^{2 \; \pi} g(\mathbf{x}_G + \vec{\rho}\,) d\alpha
\end{equation}
where $\mathbf{x}_G$ is the guiding center radial coordinate: it is related to $\mathbf{x}$, the position of the particle in the real space, through the Larmor radius $\rho$, i.e. $\mathbf{x} = \mathbf{x}_G + \vec{\rho}$, which in turn is defined as:
\[  
\vec{\rho} = \rho(\cos(\alpha) \mathbf{e}_{\perp 1} + \sin(\alpha)\mathbf{e}_{\perp 2})
\]
where $\alpha\in [0,2\pi]$ represents the gyrophase angle and $\mathbf{e}_{\perp 1}$, $\mathbf{e}_{\perp 2}$ the unit vectors of a Cartesian basis in a plane perpendicular to the magnetic field direction $\mathbf{b} = \mathbf{B}/|B|$. The function $f$ and $g$ in equation \eqref{eq:gyroaverage_definition} are defined such that $f: (r,\theta) \in \mathbb{R}^{+} \times \mathbb{R}  \mapsto f(r,\theta)$ is a polar function and $g: (x_1,x_2) \in \mathbb{R}^2 \mapsto g(x_1,x_2)$ is a Cartesian function such that $g(r \cos\theta, r \sin\theta) = f(r,\theta)$ for any pair $(r,\theta)$. The two functions represent an arbitrary field quantity respectively defined on a grid with polar and Cartesian coordinates.

\begin{figure}[!h] \centering
\includegraphics[width=0.25\textwidth,keepaspectratio]{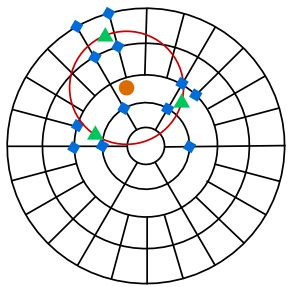}
\caption{Illustration of the procedure used to compute the gyroaverage: the function value for which we want to compute the gyroaverage is represented by ~\rondrouge, and the red circumference marks the gyroradius which has been considered. The green triangles~\trianglevert~are the points chosen to compute the gyroaverage, while the blue squares~\carrebleu~are those needed for the interpolation of the function on the green points (Lagrange interpolation of order 1).}
\label{fig:TGdescription} \end{figure}

It can be shown \cite{gyroaverage_steiner} %
that the gyroaverage operator defined in equation \eqref{eq:gyroaverage_definition} can be expressed as a function of the Bessel function of first order, and thus in the Fourier space the gyroaverage is reduced to a multiplication with a Bessel function. In this context though, another approach has been used in order to compute the gyroaverage operator, based on the 2D Lagrangian interpolation. In summary, this method consists in averaging the value of the function over $N$ points equally distributed on a circle of radius $\rho$: since these points will unlikely correspond to grid points, an interpolation method is used in order to retrieve the value of the function, according to the interpolation procedure described in Section \ref{sec:interp}. This procedure is clarified in Figure \ref{fig:TGdescription}: the function value for which we want to compute the gyroaverage is marked by an orange circle~\rondrouge, and the red circumference marks the gyroradius which has been considered. Three triangle green points~\trianglevert~are chosen to compute the gyroaverage, and since they do not correspond to any grid point, the value of the function must be retrieved with a preliminary interpolation, using the nearest grid points available, shown in figure as blue squares~\carrebleu~gathered around the triangles.

We can write the rigorous expression of the operator in the following way \cite{gyroaverage_steiner}:

\begin{equation}\label{eq:gyroaverage_definition_by_interp}
\textit{J}_\rho(f)_{j,k} \simeq \frac{1}{2\pi} \sum_{\ell=0}^{N-1} \textit{P}(f)(r_j \cos\theta_k + \rho \cos\alpha_\ell, r_j \sin\theta_k + \rho \sin\alpha_\ell) \Delta \alpha
\end{equation}
where $\alpha_\ell = \ell\Delta \alpha$, $\Delta \alpha = 2\pi/N$. 
$\textit{P}(f)$ is the Lagrange interpolator operator. Radial projection on the border of the domain is used if the points selected for the gyroaverage lie outside the domain for large radius.
%
The requirements on the gyroaverage operator are to be accurate enough in order not to disrupt the data, and to be cheap enough from a computational point of view, since it needs to be applied many times during a simulation. It is expected that the present implementation on the new grid will make the application of the gyroaverage operator cheaper and faster, with a general benefit for the global  simulation execution time.

\subsection{Advection operator}
\label{sec:advection}

Advection consists in the transport of a scalar or vectorial quantity
over a vector field. In our case, the transported quantity is the
distribution function. The advection is performed backward (Backward
Semi-Lagrangian scheme) which means that considering a grid point at
time step $t^{N+1}$ we perform the advection with a velocity field in
the opposite direction to find where the quantity was at time step
$t^N$ (see Figure~\ref{fig:advec}). As the displaced point at time
$t^N$ seldom corresponds to another grid point, a Lagrange
interpolation is performed. 

The general equation solved by the advection operator for the given distribution
function $f$ at point $(x,y)$ is:
\begin{equation}
  \label{eq:advec}
  f(x,y,t^{N+1}) = f(x-v_x\Delta t, y-v_y\Delta t, t^N)
\end{equation}
where $v_x$ and $v_y$ are the velocities along their respective dimensions and
$\Delta t$ is the time step. The right-hand side term is solved as explained
above by calling the interpolation operator described in \ref{sec:interp}. In
\gysela, velocities are defined using a Taylor expansion as described in \cite[p. 402]{grandgirard2006}.

\begin{figure}[!h] \centering 
\includegraphics[width=0.45\textwidth]{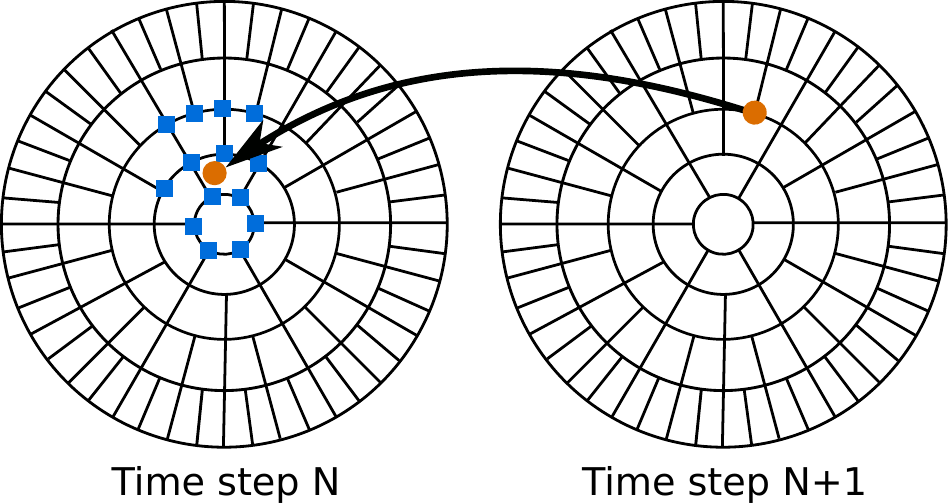}
\caption{Advection of a point~\rondrouge~of the poloidal plane and
 interpolation from time step $t^N$ points~\carrebleu~(Lagrange interpolation of
 order 3).}
\label{fig:advec}
\end{figure}



\subsection{2D finite differences for Poisson solver in polar coordinates}
\label{sec:poisson}
As said in introduction, in a gyrokinetic code the 5D Vlasov equation is coupled to a 3D quasi-neutrality equation. In \gysela{} code this last equation is projected in Fourier space in the $\theta$ dimension and solved by 1D finite differences in the radial direction. This numerical treatment is well adapted to concentric circular magnetic configuration but will be no longer applicable to more realistic magnetic configuration. Radial and poloidal directions can indeed no more be split and a 2D treatment of the poloidal $(r,\theta)$ cross-section is required. A 2D finite element method is often used in the gyrokinetic codes including D-shape magnetic configurations. For the Poisson solver, we will examine two specific meshes: (i) a non-uniform circular mesh (see Figure~\ref{fig:newgrid}) and (ii) a uniform mesh based on a large aspect ratio equilibrium and mapping (see Section~\ref{sec:mapping}). We choose to use finite differences to solve this problem. In this section, we consider the 2D Poisson equation in polar coordinates on a domain $\Omega$,
\begin{equation}
  \label{eq:poiss_polar}
  \frac{\partial^2 \ff}{\partial r^2}+\frac{1}{r}\frac{\partial \ff}{\partial r}+\frac{1}{r^2}\frac{\partial^2 \ff}{\partial \theta^2}=R(r,\theta)
\end{equation}
with Dirichlet boundary conditions $\ff(r=\rmax,\theta)=g(\theta)$ on $\partial\Omega$. 

\subsubsection{2D finite differences for a non-uniform circular mesh} 
\begin{figure}[h]
  \centering
  \begin{tabular}{cc}
    \includegraphics[width=0.55\linewidth]{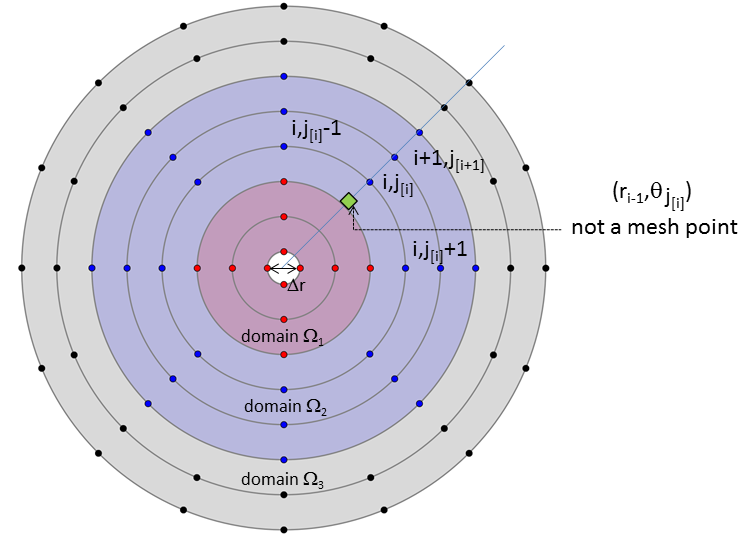} &
    \includegraphics[width=0.45\linewidth]{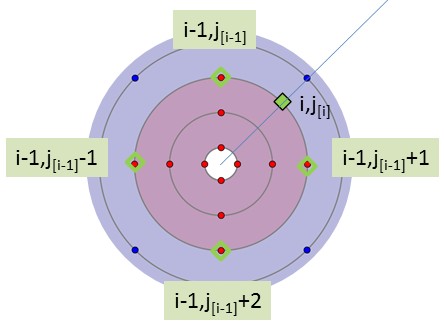} \\
    (a) & (b)
  \end{tabular}
  \caption{(a) Example of a non-uniform mesh divided into three parts with 4 points by radius in the first one, 8 in the second one and 16 points for the last one. To compute value at mesh point $(r_i,\theta_j)$ with finite differences of second order the points $(r_i,\theta_{j_{[i]}+1})$, $(r_i,\theta_{j_{[i]}-1})$, $(r_{i+1},\theta_{j_{[i]}})$ and $(r_{i-1},\theta_{j_{[i]}})$. The problem is that the last one $(r_{i-1},\theta_{j_{[i]}})$ is not a mesh point. (b) The value at position $(r_{i-1},\theta_{j_{[i]}})$ is computed by Lagrange interpolation of 3rd order by using the mesh points represented by a square.}
  \label{fig:non_uniform_mesh}
\end{figure}

Let us first consider equation \eqref{eq:poiss_polar} on a disk $\Omega=\{(r,\theta):0<r<\rmax \text{ with } \rmax\in \mathbb{R} \text{ and } 0\le \theta \le 2\pi \}$ where $\Omega$ is described by a non-uniform circular mesh $\Omega_k$.
To overcome the singularity problem at $r=0$, we use the same centered finite difference method as proposed in Lai's paper~\cite{lai2001note}
\footnote{Note that another trick to cope with the singularity at the origin can be found in  \cite{strikwerda2004finite}, p334,
see also \cite{wright2013fourth} and references therein for further references. In particular, the idea to use a shifted mesh of one half mesh size seems to date back to \cite{eisen1967numerical}.}.
One of the trick consists in solving Equation \eqref{eq:poiss_polar} for $r\in [\rmin,\rmax]$ with $\rmin=\Delta r/2$ and a half-integered grid in radial direction and an integered grid in poloidal direction.
In this section, we propose an extension of the method proposed for an uniform circular mesh by Lai to a non-uniform one.
The difficulty is to adapt the scheme to allow a different number of poloidal points per radius.
This implies the adding of interpolations.
The scheme proposed in the following is based on Lagrange interpolation of third order which is a good compromise between accuracy and complexity.
Let $N=N_r-1$ be the number of cells in radial direction and $\Nthetai{i}$ be the number of cells along $\theta$ on the circle of radius $r_i$. 
Let us call $\gamma_i$ the ratio between number of poloidal mesh points for $r=r_i$ and the one for circle of radius $r=r_{i-1}$, namely $\gamma_i = \Nthetai{i}/\Nthetai{i-1}$. Let us add the two constraints on $\Omega_k$: (i) $\Nthetai{i}$ is even and (ii) $\gamma_i\ge 1$.
Then, $\Omega_k$ is defined as 
\begin{align}
  r_i              &= (i-\frac{1}{2})\Delta r \quad \text{for all }i=1,2,\cdots,N+1 \label{eq:r_grid}\\
  \theta_{j_{[i]}} &= (j_{[i]}-1)\Dthetai{i} \quad \text{for all }j_{[i]}=1,2,\cdots,\Nthetai{i}+1 \label{eq:theta_grid}
\end{align} 
where $\Delta r = \rmax/(N+1/2)$ and $\Dthetai{i}=2\pi/\Nthetai{i}$. 
Let us notice that these indexes used are different from the one used in section \ref{sec:non_uniform} (indices starting here at 1 instead of 0 previously).
Let the discrete values be denoted by $\Rij=R(r_i,\theta_j)$, $g_j=g(\theta_j)$ and $\ffij=\ff_{i,j_{[i]}}$ where $\ff_{i,j_{[k]}}=\ff(r_i,\theta_{j_{[k]}})$.
Then, the discrete version of Eq.~\eqref{eq:poiss_polar} becomes, for $i=1,\cdots,N$ and $j_{[i]}=1,2,\cdots,\Nthetai{i}$:
\begin{align}
  \label{eq:poiss_polar_discrete_nonunif}
  \frac{\ff(r_{i+1},\theta_{j_{[i]}})-2\ffij+\fftilde(r_{i-1},\theta_{j_{[i]}})}{(\Delta r)^2} &+ 
  \frac{1}{r_i} \frac{\ff(r_{i+1},\theta_{j_{[i]}})-\fftilde(r_{i-1},\theta_{j_{[i]}})}{2\Delta r} \nonumber\\
  &\qquad\qquad+\frac{1}{r_i^2}\frac{\ffi{j+1}-2\ffij+\ffi{j-1}}{(\Dthetai{i})^2} = R_{i,j}
\end{align}  
where the boundary values are given: (i) radially by the Dirichlet condition $\ff_{N+1,j}=g_j$ for all $j\in [1,\Nthetai{N+1}]$ and (ii) poloidally by $\ff_{i,0}=\ff_{i,\Nthetai{i}}$ and $\ff_{i,1}=\ff_{i,\Nthetai{i}+1}$ for all $i\in [1,N+1]$ due to $2\pi$ periodic boundary conditions.
The term $\ff(r_{i+1},\theta_{j_{[i]}})$ is equal to $\ff(r_{i+1},\theta_{j_{[i+1]}})$ with $j_{[i+1]}=(j_{[i]}-1)\gamma_{i+1}+1$ where $(r_{i+1},\theta_{j_{[i+1]}})$ is a mesh point, so $\ff(r_{i+1},\theta_{j_{[i]}})=\ff_{i+1,(j_{[i]}-1)\times\gamma_{i+1}+1}$.
The term $\fftilde(r_{i-1},\theta_{j_{[i]}})$ corresponds to an approximation of $\ff$ at the point $(r_{i-1},\theta_{j_{[i]}})$ because if $\gamma_i>1$ then $\theta_{j_{[i]}}$ is not automatically a mesh point (see Figure \ref{fig:non_uniform_mesh}). The value $\fftilde(r_{i-1},\theta_{j_{[i]}})$ is defined as
\begin{equation}
  \fftilde(r_{i-1},\theta_{j_{[i]}}) = \left\{
  \begin{array}{l}
    \ff(r_{i-1},\theta_{j_{[i-1]}}) \text{ if }j_{[i-1]}=(j_{[i]}-1)/\gamma_i+1 \in \NN \\
    \text{approximation of }f \text{ at position }(r_{i-1},\theta_{j_{[i]}}) \text{ otherwise}
  \end{array}
  \right.
\end{equation}
The required approximations are computed by using a Lagrange interpolation of third order. So let us consider $k\in \mathbb{N}$ the integer such that $\theta_k<\theta_{j_{[i-1]}}<\theta_{k+1}$, then using \eqref{eq:lagrange_poly}-\eqref{eq:coeff_lagrange} notations,
\begin{equation}
  \fftilde(r_{i-1},\theta_{j_{[i]}}) \approx \sum_{n=-1}^{2} L^{(3)}_n(\theta_{j_{[i-1]}})\ff(r_{i-1},\theta_{k+n}) \quad\text{with }j_{[i-1]}={\rm int}\left(\frac{j_{[i]}-1}{\gamma_i}\right)+1
\end{equation} 
where the Lagrange polynomials $L^{(3)}_n$ are defined by
\begin{equation}
  \label{eq:Lagrange_polynom_order3}
  L^{(3)}_n(\theta)=\prod_{\substack{i=-1 \\ i\ne n}}^2 \frac{(\theta-\theta_{k+i})}{ (\theta_{k+n}-\theta_{k+i})} 
  \quad\text{for all } \theta\in[\theta_k,\theta_{k+1}[\quad\text{and }n=-1,\cdots,2
\end{equation}

Let us define, for all $i\in [1,N+1]$,
\begin{equation}
  \label{eq:lambdai_betai}
  \lambda_i=\frac{\Delta r}{2r_i}=\frac{1}{2(i-1/2)}
  \quad\text{and}\quad
  \beta_i = \frac{(\Delta r)^2}{r_i^2(\Dthetai{i})^2}=\frac{1}{(i-1/2)^2(\Dthetai{i})^2}
\end{equation}
Then, equation \eqref{eq:poiss_polar_discrete_nonunif} reads
\begin{align}
  & (1-\delta_{i,1})(1-\lambda_i)\fftilde(r_{i-1},\theta_{j_{[i]}}) + \beta_i\left[(1-\delta_{j,1})\ff_{i,j-1} + \delta_{j,1}\ff_{i,\Nthetai{i}}\right] \nonumber\\   
  & \qquad\qquad - (2+2\beta_i)\ff_{i,j} + \beta_i\left[(1-\delta_{j,\Nthetai{i}})\ff_{i,j+1}+\delta_{j,\Nthetai{i}}\ff_{i,1}\right] \nonumber\\
  & \qquad\qquad + (1-\delta_{i,N})(1+\lambda_i)\ff_{i+1,(j_{[i]}-1)\gamma_{i+1}+1}
  = (\Delta r)^2 R_{i,j} - \delta_{i,N}(1+\lambda_{N})g_j \label{eq:poiss_polar_discrete_nonunif_1}
\end{align}
Let us notice that due to the choice of $r_1=\Delta r/2$, $(1-\lambda_1)=0$.
At the opposite of what is proposed in Lai's paper, let us order the unknowns $\ff_{i,j}$ radius by radius, such that the unknown vector $u$ of size $N_{\rm tot}=\displaystyle\sum_1^{N} \Nthetai{i}$ is defined as
\begin{equation}
  \label{eq:unknown_vector_nonunif}
  u^t=\left[\ff_1\;,\ff_2\;,\cdots,\ff_{N}\right] \quad\text{with}\quad \ff_i^t = \left[\ff_{i,1}\;,\ff_{i,2}\;,\cdots,\ff_{i,\Nthetai{i}}\right]
\end{equation}
The matrix system associated to the discrete equation system \eqref{eq:poiss_polar_discrete_nonunif_1} reads $Au=b$ where $A$ is a $N_{\rm tot}\times N_{\rm tot}$ sparse matrix, given by
\begin{equation}
\label{eq:A_matrix_nonunif}
A = 
\begin{bmatrix}
	A_1          & \Lambda_1^+   &               &            &                     &                         &                   \\ 
	\TLambda_2^- &  A_2          &  \Lambda_2^+  &            &                     &           0             &                   \\ 
	             &  \ddots       &  \ddots       &  \ddots    &                     &                         &                   \\ 
	             &               & \TLambda_i^-  &     A_i    &  \Lambda_i^+        &                         &                   \\ 
		     &               &               &  \ddots    &  \ddots             &  \hspace{-0.5cm}\ddots  &                   \\
		     &          0    &               &            &  \TLambda_{N-1}^- &  A_{N-1}              & \Lambda_{N-1}^+ \\ 
	             &               &               &            &                     & \TLambda_{N}^-        & A_{N} 
\end{bmatrix} 
\end{equation}
The $N$ matrices $A_i$ are $\Nthetai{i}\times \Nthetai{i}$ matrix defined as
\begin{equation}
  \label{eq:Ai_nonunif}
  A_i =
  \begin{bmatrix}
  -(2+2\beta_i)  & \beta_i       &    0        &             &  \beta_i      \\ 
  \beta_i        & -(2+2\beta_i) &  \beta_i    &                             \\ 
                 &               &  \ddots     &   \ddots    &               \\ 
                 &               &  \ddots     &   \ddots    & \beta_i       \\ 
   \beta_i       &    0          &             &  \beta_i    & -(2+2\beta_i) 
  \end{bmatrix}
\end{equation}
The matrix $\Lambda_i^+$ is a $\Nthetai{i}\times \Nthetai{i+1}$ matrix while $\TLambda_i^-$ is a $\Nthetai{i}\times \Nthetai{i-1}$ matrix, both defined as
\begin{equation}
  \Lambda_i^+ = 
  \begin{bmatrix}
  D_i^1  \\ 
  \vdots \\
  D_i^k \\ 
  \vdots \\
  D_i^{\Nthetai{i}}
  \end{bmatrix} 
  \qquad\text{and}\qquad
  \TLambda_i^- = 
  \begin{bmatrix}
  C_i^1  \\ 
  \vdots \\
  C_i^k \\ 
  \vdots \\
  C_i^{\Nthetai{i-1}}
  \end{bmatrix}   
\end{equation}
where $D_i^k$ is a line matrix $1\times\Nthetai{i+1}$ matrix where all elements are equal to $0$ except the $((k-1)\gamma_{i+1}+1)$th term is equal to $(1+\lambda_i)$.
Furthermore, $C_i^k$ is the $\gamma_i\times\Nthetai{i-1}$ matrix given by 
\begin{equation*}
  C_i^k = (1-\lambda_i)
  \begin{bmatrix}
    \;0 &           &  \cdots &            0                 &          1                 &         0                  &         \cdots             &   &         & 0 \\ 
        & \cdots    &    0    &  L_{-1}(\kappa_i^k+1)        & L_{0}(\kappa_i^k+1)        & L_{1}(\kappa_i^k+1)        & L_{2}(\kappa_i^k+1)        & 0 &  \cdots &   \\ 
        &           &    0    &       \vdots                 &       \vdots               &    \vdots                  &      \vdots                & 0 &         &   \vspace{0.2cm}\\ 
      0 & \cdots    &    0    &  L_{-1}(\kappa_i^k+\gamma_i) & L_{0}(\kappa_i^k+\gamma_i) & L_{1}(\kappa_i^k+\gamma_i) & L_{2}(\kappa_i^k+\gamma_i) & 0 &  \cdots &  0\;
  \end{bmatrix} 
\end{equation*}
with $\kappa_i^k=(k-1)\gamma_i+1$ and $L_m$ the Lagrange polynomials defined by Eq.\eqref{eq:Lagrange_polynom_order3} where for more readability, $L_m(l)=L_m^{}(\theta_{j_{[i]}=l})$. Let us notice that the column position of the value $1$ of first row of $C_i^k$ is equal to $k$.
Finally, the right hand side vector $b$ can be expressed as
\begin{equation}
\label{eq:b_nonunif}
b = 
\begin{bmatrix}
b_1\\ 
b_2\\ 
\vdots\\ 
b_{N}
\end{bmatrix}
\quad\text{with}\quad
b_i = (\Delta r)^2 
\begin{bmatrix}
R_{i,1}\\ 
R_{i,2}\\ 
\vdots\\ 
R_{i,\Nthetai{i}}
\end{bmatrix}
-\delta_{i,N}(1+\lambda_{N})
\begin{bmatrix}
g_1\\ 
g_2\\ 
\vdots\\ 
g_{N}
\end{bmatrix}
\end{equation}
Let us notice, that Poisson equation \eqref{eq:poiss_polar} on a circular uniform mesh ($\Nthetai{i}=N_\theta, \,\forall i$) can be trivially deduced from the previous $Au=b$ matrix system. The $N_{\rm tot}\times N_{\rm tot}$ matrix $A$ is given by Eq. \eqref{eq:A_matrix_nonunif} with $N_{\rm tot}=N\,\Ntheta$ where $\TLambda_i^-$ and $\TLambda_i^+$ are $\Ntheta\times \Ntheta$ diagonal matrices with ${\rm diag}(\TLambda_i^\pm) = 1\pm\lambda_i$. 

\subsubsection{2D finite differences on a mapped uniform mesh}
One difficulty was to extend the Poisson solver \cite{lai2001note} to a non uniform mesh, as done in the previous subsection. Another one is to deal with a mapping. So, we focus here on this point, starting with
a uniform mesh. The combination of both schemes will be the subject of further work and is not tackled here.
 We refer to  \cite{hamiaz2015semi} for the use of a Mudpack solver, and \cite{HamiazBack} for the use of a finite element solver based on B-splines.
Such solvers might be adapted, but here we consider a specific treatment for the center; so we develop a stand-alone solution with finite differences. Note that some adaptations have to be done with respect to the previous case \cite{lai2001note} and we will propose two examples of solvers with $7$ and $9$ points (we could not get a $5$ points solution, here due to the appearance of mixed terms from the mapping as we will see).
We consider here the Poisson equation first on a elliptic domain and then for the large aspect ratio mapping (see Eq.~\eqref{equation_culham} and~\eqref{param_culham} for the latter).
We write
$$
x(r,\theta)=ar\cos(\theta),\ y(r,\theta)=br\sin(\theta).
$$
From Eq.\eqref{eq:poiss_polar}, which reads $\Delta U=F$ and writing $u(r,\theta)=U(x(r,\theta),y(r,\theta)),\ f(r,\theta)=F(x(r,\theta),y(r,\theta))$, we have the relations
$$
\nabla_{x,y}U=J^{-T}\nabla_{r,\theta}u,\ \nabla_{x,y}\cdot A=\frac{1}{|J|}\nabla_{r,\theta}\cdot(|J|J^{-1}A),\ J=\left(\begin{array}{ll}
 \frac{\partial x}{\partial r} &  \frac{\partial x}{\partial \theta}\\ 
 \frac{\partial y}{\partial r}& \frac{\partial y}{\partial \theta}\\ 
\end{array}\right),\ |J| =\det(J),
$$
which lead to
$$
\Delta U = \nabla_{x,y}\cdot \nabla_{x,y}U = \frac{1}{|J|}\nabla_{r,\theta}\cdot \left(|J|G\nabla_{r,\theta}u\right),\ G = J^{-1}J^{-T}.
$$
We have here
$$
J^{-T}=\frac{1}{|J|}\left(\begin{array}{ll}
\frac{\partial y}{\partial \theta} &-\frac{\partial y}{\partial r} \\ 
 -\frac{\partial x}{\partial \theta} & \frac{\partial x}{\partial r} \\ 
\end{array}\right),\ 
|J|G=\frac{1}{|J|}\left(\begin{array}{ll}
(\frac{\partial x}{\partial \theta})^2+(\frac{\partial y}{\partial \theta})^2 &-\frac{\partial x}{\partial r}\frac{\partial x}{\partial \theta}-\frac{\partial y}{\partial r}\frac{\partial y}{\partial \theta}\\
-\frac{\partial x}{\partial r}\frac{\partial x}{\partial \theta}-\frac{\partial y}{\partial r}\frac{\partial y}{\partial \theta}& (\frac{\partial x}{\partial r})^2+(\frac{\partial y}{\partial r})^2
\end{array}\right).
$$

\noindent  For an ellipse, we have $|J| = abr$ and
$$
J^{-T}=\frac{1}{|J|}\left(\begin{array}{ll}
br\cos(\theta) & -b\sin(\theta)\\ 
 ar\sin(\theta)& a\cos(\theta)\\ 
\end{array}\right),\  |J|G=\frac{1}{|J|}
\left(\begin{array}{ll}
r^2\left(b^2\cos^2(\theta)+a^2\sin^2(\theta)\right) & (a^2-b^2)r\sin(\theta)\cos(\theta)\\ 
(a^2-b^2)r\sin(\theta)\cos(\theta)&a^2\cos^2(\theta)+b^2\sin^2(\theta)\\ 
\end{array}\right)
$$

\noindent For the large aspect ratio mapping, we have $|J|=(1-E_0)br-2\delta_0br^2\cos(\theta)$, $$
J^{-T}=\frac{1}{|J|}\left(\begin{array}{ll}
(1+E_0)r\cos(\theta) &-(1+E_0)\sin(\theta) \\ 
 (1-E_0)r\sin(\theta) &(1-E_0)\cos(\theta)-2\delta_0r \\ 
\end{array}\right)$$ and
$${\small
|J|G=\frac{1}{|J|}\left(\begin{array}{ll}
r^2((1+E_0)^2\cos^2(\theta)+(1-E_0)^2\sin^2(\theta))& -4E_0r\cos(\theta)\sin(\theta)-2(1-E_0)\delta_0r^2\sin(\theta)\\
-4E_0r\cos(\theta)\sin(\theta)-2(1-E_0)\delta_0r^2\sin(\theta) & (1+E_0)^2\sin^2(\theta)+((1-E_0)\cos(\theta)-2\delta_0r)^2 \\
\end{array}\right).}
$$

\noindent Writing $|J|G =(a_{ij})$, we get the equation
$$
\frac{\partial}{\partial r}\left(a_{11}\frac{\partial u}{\partial r}+a_{12}\frac{\partial u}{\partial \theta}\right)+\frac{\partial}{\partial \theta}\left(a_{21}\frac{\partial u}{\partial r}+a_{22}\frac{\partial u}{\partial \theta}\right)=|J|f,\ 0<r<1,\ 0\le \theta< 2\pi
$$
and $u(1,\theta) = g(\theta)$.

\noindent Let  $N,M \in \mathbb{N}^*$. We write
$$
r_i=(i-1/2)\Delta r,\ \Delta r=\frac{2}{2N+1},\ i=1,\dots,N+1; \ \theta_j = (j-1)\Delta \theta,\ \Delta \theta = \frac{2\pi}{M},\ j=1,\dots,M+1.
$$

\noindent We consider the following finite difference scheme with $7$ points
\begin{eqnarray*}
&&\frac{1}{\Delta r^2}\left(a_{11}^{i+1/2,j}(u_{i+1,j}-u_{ij})-a_{11}^{i-1/2,j}(u_{i,j}-u_{i-1,j})\right)\\
&&+\frac{1}{2\Delta r\Delta \theta}\left(a_{12}^{i+1/2,j}(u_{i,j+1}-u_{ij})-a_{12}^{i-1/2,j}(u_{i-1,j+1}-u_{i-1,j})\right)\\
&&+\frac{1}{2\Delta r\Delta \theta}\left(a_{12}^{i+1/2,j}(u_{i+1,j}-u_{i+1,j-1})-a_{12}^{i-1/2,j}(u_{i,j}-u_{i,j-1})\right)\\
&&+\frac{1}{2\Delta r\Delta \theta}\left(a_{21}^{i,j+1/2}(u_{i+1,j}-u_{ij})-a_{21}^{i,j-1/2}(u_{i+1,j-1}-u_{i,j-1})\right)\\
&&+\frac{1}{2\Delta r\Delta \theta}\left(a_{21}^{i,j+1/2}(u_{i,j+1}-u_{i-1,j+1})-a_{21}^{i,j-1/2}(u_{i,j}-u_{i-1,j})\right)\\
&&+\frac{1}{\Delta \theta^2}\left(a_{22}^{i,j+1/2}(u_{i,j+1}-u_{ij})-a_{22}^{i,j-1/2}(u_{i,j}-u_{i,j-1})\right)=|J|_{ij}f_{ij},
\end{eqnarray*}
for  $j=1,\dots,M$ and $i=1,\dots,N$.
We have here $f_{ij}=f(r_i,\theta_j)$, $|J_{ij}|=J(r_i,\theta_j)$ and $a_{k\ell}^{pq}=a_{k\ell}(r_p,\theta_q),\ p,q\in \frac{1}{2}\mathbb{Z}$.
\noindent The system is modified as follows in order to deal with the boundary conditions
\begin{itemize}
\item $u_{ij}$ is replaced by $u_{i\tilde{j}}$ where $j=\tilde{j}+kM, k\in \mathbb{Z}$ and $1\le \tilde{j}\le M$. 
\item $u_{0,j}$ is replaced by $u_{1,M/2+j}$ for $j=1,\dots,M/2$
and by $u_{1,M/2-j}$ for $j=M/2+1,\dots,M$.
\item $u_{N+1,j}$ is replaced by $g_j=g(\theta_j)$, for $j=1,\dots,M$.
\end{itemize}

\noindent Note that here a $7$ points stencil is needed. 
\noindent We have to take special care on the boundary condition: $u_{0,j}$ does not cancel and it is replaced by $u_{1,N/2-j}$ (we assume here that $N$ is even). 
\noindent For the case of a circle, we get the standard $5$ points stencil (the terms $a_{21},a_{2,2}$ cancel) and $u_{0,j}$ cancels.
\noindent Next, we give also another scheme with a $9$ points stencil, that is using contributions of $u_{i-1,j-1}$ and $u_{i+1,j+1}$.
\noindent With respect to \cite{lai2001note}, we underline that the following adaptations have been done:
\begin{itemize}
\item a $5$ points stencil is (at least seems) no more possible for second order accuracy because of the mixed terms, and several schemes are possible
(see \cite{jovanovic2013analysis} p204, \cite{hackbush2017} p103 and \cite{strikwerda2004finite} p335 for similar schemes using a stencil with $7$ or $9$ points)
\item due to the mixed terms,  the term $u_{0,j}$ does not cancel, we have to use the value $u_{1,N/2-j}$, as done in \cite{lai2002simple}.
\end{itemize}
\paragraph{A $9$ points scheme}
\noindent We now derive another scheme for the mixed terms $a_{k,\ell}=a_{21}^{k,\ell}=a_{12}^{k,\ell}$.
For $\frac{\partial}{\partial r}\frac{\partial au}{\partial \theta}$, we can use $\frac{1}{4}a_{i-1/2,j+1/2}(u_{i,j+1}+u_{i,j}+u_{i-1,j+1}+u_{i-1,j})$ and
$\frac{1}{4}a_{i-1/2,j-1/2}(u_{i,j}+u_{i,j-1}+u_{i-1,j}+u_{i-1,j-1})$, which gives
$$
\frac{1}{4\Delta \theta}a_{i-1/2,j-1/2}(u_{i,j}+u_{i,j-1}+u_{i-1,j}+u_{i-1,j-1})-\frac{1}{4\Delta \theta}a_{i-1/2,j+1/2}(u_{i,j+1}+u_{i,j}+u_{i-1,j+1}+u_{i-1,j}),
$$
and
\begin{eqnarray*}
4\Delta r\Delta \theta\frac{\partial}{\partial r}\frac{\partial au}{\partial \theta} &&
\simeq-a_{i+1/2,j-1/2}(u_{i+1,j}+u_{i+1,j-1}+u_{i,j}+u_{i,j-1})\\
&&+a_{i+1/2,j+1/2}(u_{i+1,j+1}+u_{i+1,j}+u_{i,j+1}+u_{i,j})\\
&&+a_{i-1/2,j-1/2}(u_{i,j}+u_{i,j-1}+u_{i-1,j}+u_{i-1,j-1})\\
&&-a_{i-1/2,j+1/2}(u_{i,j+1}+u_{i,j}+u_{i-1,j+1}+u_{i-1,j}).
\end{eqnarray*}

\noindent We have also
$$
\frac{\partial}{\partial r}a\frac{\partial u}{\partial \theta}+\frac{\partial}{\partial \theta}a\frac{\partial u}{\partial r}
=\frac{\partial}{\partial r}\frac{\partial au}{\partial \theta}-\frac{\partial}{\partial r}\frac{\partial a}{\partial \theta}u+
\frac{\partial}{\partial r}\frac{\partial u}{\partial \theta}a.
$$

\noindent We finally get
\footnote{We have the intermediate steps:
\begin{eqnarray*}
&&4\Delta r\Delta \theta\left(\frac{\partial}{\partial r}a\frac{\partial u}{\partial \theta}+\frac{\partial}{\partial \theta}a\frac{\partial u}{\partial r}\right)\\
&&=-a_{i+1/2,j-1/2}(u_{i+1,j}+u_{i+1,j-1}+u_{i,j}+u_{i,j-1})+a_{i+1/2,j+1/2}(u_{i+1,j+1}+u_{i+1,j}+u_{i,j+1}+u_{i,j})\\
&&+a_{i-1/2,j-1/2}(u_{i,j}+u_{i,j-1}+u_{i-1,j}+u_{i-1,j-1})-a_{i-1/2,j+1/2}(u_{i,j+1}+u_{i,j}+u_{i-1,j+1}+u_{i-1,j})\\
&&+a_{i,j}(u_{i+1,j+1}+u_{i-1,j-1}-u_{i-1,j+1}-u_{i+1,j-1})\\
&&-4\left(a_{i+1/2,j+1/2}+a_{i-1/2,j-1/2}-a_{i-1/2,j+1/2}-a_{i+1/2,j-1/2}\right)u_{i,j}\\
&&=-a_{i+1/2,j-1/2}(u_{i+1,j}+u_{i+1,j-1}+u_{i,j-1})+a_{i+1/2,j+1/2}(u_{i+1,j+1}+u_{i+1,j}+u_{i,j+1})\\
&&+a_{i-1/2,j-1/2}(u_{i,j-1}+u_{i-1,j}+u_{i-1,j-1})-a_{i-1/2,j+1/2}(u_{i,j+1}+u_{i-1,j+1}+u_{i-1,j})\\
&&+a_{i,j}(u_{i+1,j+1}+u_{i-1,j-1}-u_{i-1,j+1}-u_{i+1,j-1})\\
&&-3\left(a_{i+1/2,j+1/2}+a_{i-1/2,j-1/2}-a_{i-1/2,j+1/2}-a_{i+1/2,j-1/2}\right)u_{i,j}\\
\end{eqnarray*}
}

\begin{eqnarray*}
&&4\Delta r\Delta \theta\left(\frac{\partial}{\partial r}a\frac{\partial u}{\partial \theta}+\frac{\partial}{\partial \theta}a\frac{\partial u}{\partial r}\right)\\
&&=(a_{i,j}+a_{i+1/2,j+1/2})u_{i+1,j+1}+(a_{i,j}+a_{i-1/2,j-1/2})u_{i-1,j-1}\\
&&-(a_{i,j}+a_{i-1/2,j+1/2})u_{i-1,j+1}-(a_{i,j}+a_{i+1/2,j-1/2})u_{i+1,j-1}\\
&&+(a_{i+1/2,j+1/2}-a_{i+1/2,j-1/2})u_{i+1,j}-(a_{i-1/2,j+1/2}-a_{i-1/2,j-1/2})u_{i-1,j}\\
&&+(a_{i+1/2,j+1/2}-a_{i-1/2,j+1/2})u_{i,j+1}-(a_{i+1/2,j-1/2}-a_{i-1/2,j-1/2})u_{i,j-1}\\
&&-3\left(a_{i+1/2,j+1/2}+a_{i-1/2,j-1/2}-a_{i-1/2,j+1/2}-a_{i+1/2,j-1/2}\right)u_{i,j}.
\end{eqnarray*}




\section{Convergence results}
\label{sec:results}
\subsection{Interpolation}
\label{sec:interp_res}
The interpolation operator is of utmost importance, it is used as a 
building block by more complex
operators. As such, it is essential that this operator remains accurate
enough to keep the simulated physics valid. Performance are not detailed in the
paper though it is critical and impact almost every piece of the code. It will
be presented in future work where it will be integrated in \gysela\ and compared to
previous schemes. The accuracy of the Lagrange interpolation depends on
three parameters: the degree of the Lagrange polynomial, the mesh discretization
in the $r$ direction and in the $\theta$ direction. 

In order to test our implementation, we perform interpolations from the polar
mesh to a uniform Cartesian grid of size $\textstyle [-r_{max}:r_{max}]\times{}[-r_{max}:r_{max}]$
with $2\,N_r$ points in each direction. Points outside of the polar mesh are
discarded. The mesh is initialized using a sine product
$f(x,y)=\sin(5x)\times\cos(4y)$ for the following tests. Solution
is thus analytically known everywhere on the plane. The following figures give
different norms ($L_1$, $L_2$ and $L_{inf}$) of the error done when performing
the interpolation on the whole Cartesian grid. The results are given for the
uniform and non-uniform mesh.  The base mesh used in the simulation is:
$N_r = 256$, $N_\theta = 256$ and Lagrange order is 7 for the uniform mesh. For
the non-uniform mesh the Lagrange order is also 7, $N_r = 256$ and the
$N_{\theta_{[i]}}$ are given as such
$(2:\underline{32},8:\underline{64},64:\underline{128},182:\underline{256})$,
which reads: there are 32 points in $\theta$ direction on the 2 inner most
\textit{radii} ($\Nthetai{0}=\Nthetai{1}=32$), 64 points on the 8 following
\textit{radii}, and so on. This gives 15\% less points for the non-uniform mesh
than for the uniform one with $\Ntheta=256$.

\begin{figure}[!h]
  \centering
  \includegraphics[width=0.7\textwidth]{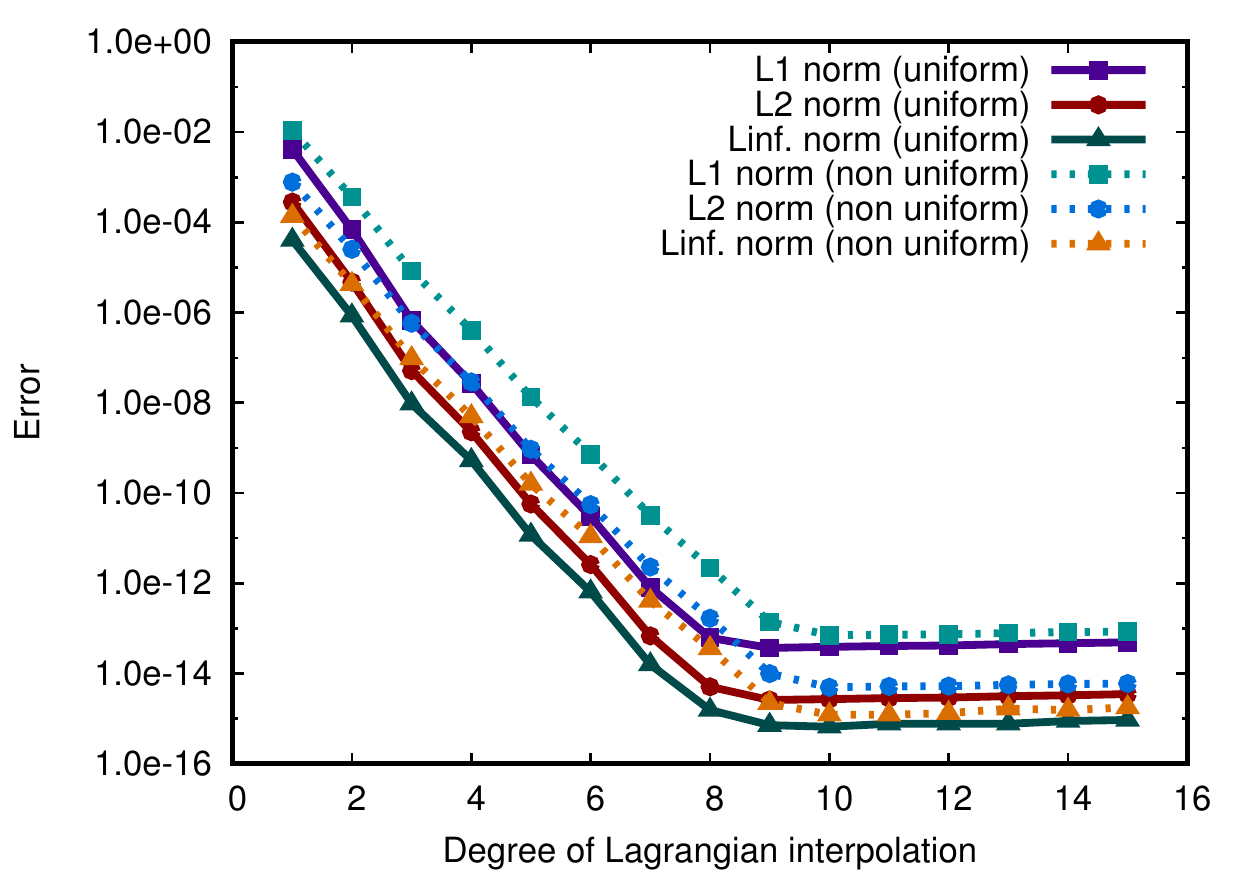}
  \caption{Convergence in degree for the 2D Lagrangian interpolation.}
  \label{fig:interp_degree}
\end{figure}

In Figure \ref{fig:interp_degree}, the error is presented against the degree of
the Lagrange interpolation which ranges from 1 to 15. Both the uniform and the
non-uniform meshes have the same behaviour. There is a convergence phase where
the error decreases steadily before stopping at a plateau. It either reaches
hardware precision or the accuracy allowed by the meshing on both dimensions. In
this case it is hardware precision for a double-precision floating-point
($10^{-15}$) which is achieved by $L_{inf}$ norm (maximum value of the
error). The non-uniform mesh proves to be less accurate because the set of
values of $N_{\theta_{[i]}}$ we have chosen is good but not optimal.

\begin{figure}[!h]
  \centering
  \begin{subfigure}[t]{0.45\textwidth}
    \centering
    \hspace*{-35pt}
    \includegraphics[width=1.2\textwidth]{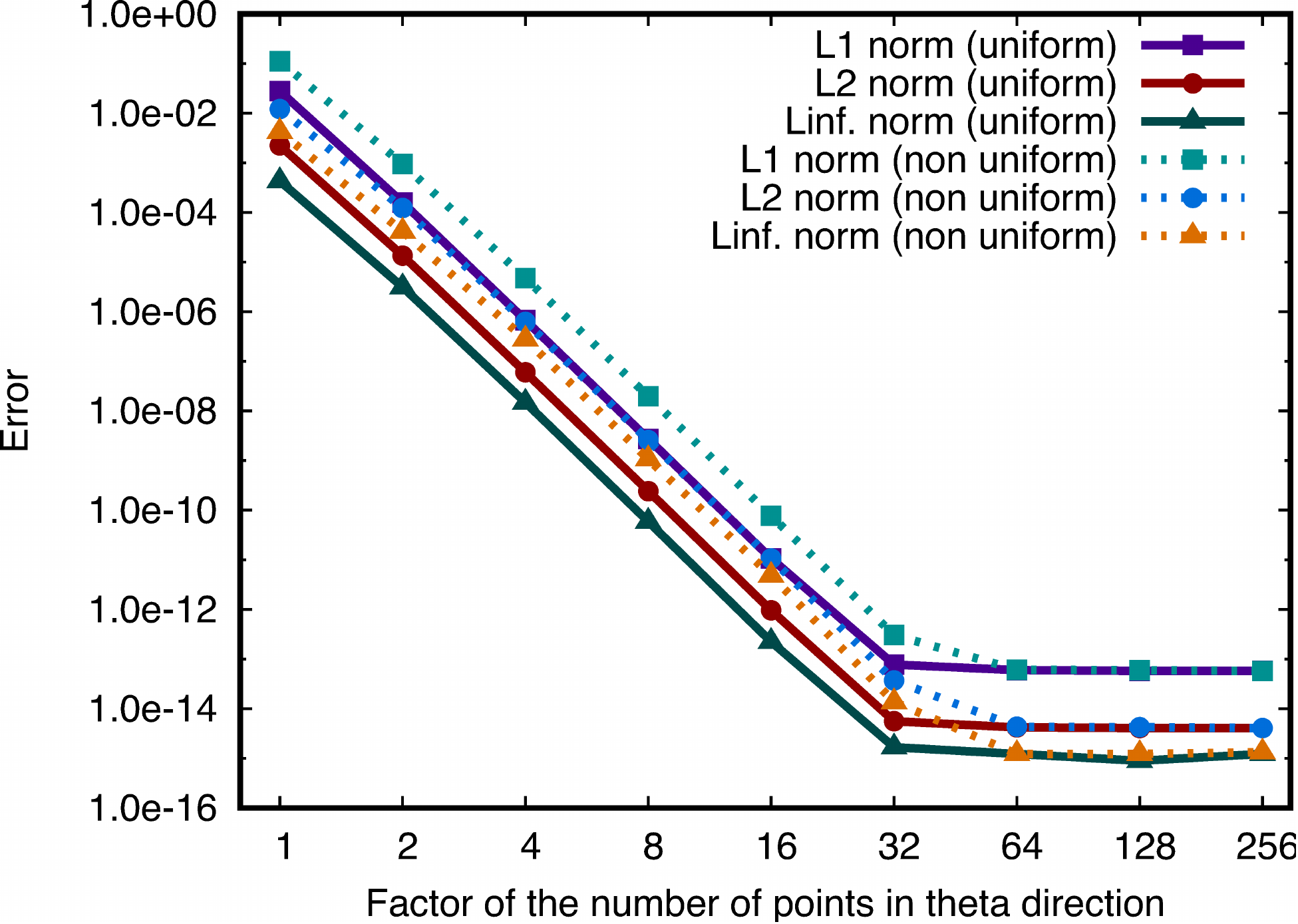}
    \caption{Convergence in $\theta$ dimension.}
    \label{fig:interp_theta}
  \end{subfigure}
  \qquad
  \begin{subfigure}[t]{0.45\textwidth}
    \centering
    \hspace*{-10pt}
    \includegraphics[width=1.2\textwidth]{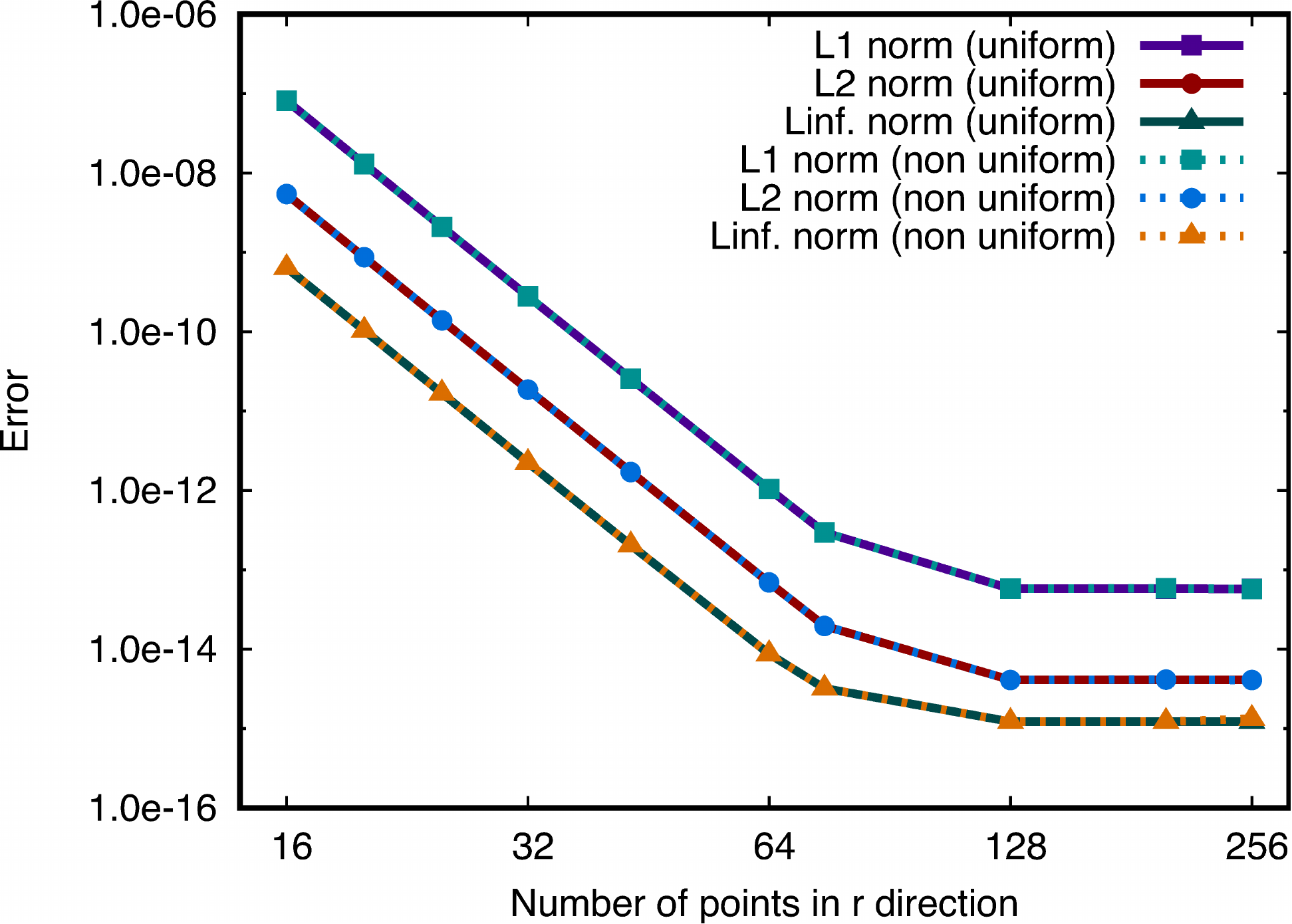}
    \caption{Convergence in $r$ dimension.}
    \label{fig:interp_r}
  \end{subfigure}
  \caption{Space convergence for the 2D Lagrangian interpolation.}
  \label{fig:space_conv}
\end{figure}

Figure~\ref{fig:interp_theta}, representing the convergence in $\theta$
direction, shows the error against the discretization factor along $\theta$. For
the uniform mesh we start at 1 with a mesh of size of $N_r=256,\, N_\theta=16$,
then 2 with $N_r=256,\, N_\theta=32$ and then at $n$ with $N_r=256,\,
N_\theta=2^{n-1}\times 16$. For the non-uniform mesh at a general discretization
factor $n$
we have $N_r=256$ and
$(2:\underline{2^{n-1}\times 2},8:\underline{2^{n-1}\times 4},
64:\underline{2^{n-1}\times 8}, 182:\underline{2^{n-1}\times 16})$. On the
convergence study in $\theta$ direction we find the same difference between the
two meshes as shown in Figure~\ref{fig:interp_degree}. The convergence rate is
the same which assesses the correctness of the interpolation operator. On Figure~\ref{fig:interp_r} the curves perfectly match because the number of points in
$\theta$ direction is chosen high enough not to influence accuracy (2048 on uniform mesh and ranged from
256 at the center to 2048 at the outter edge on non-uniform mesh). As the meshing method along $r$ has not been changed, both mesh types
gives the same convergence results.

On these simple test functions (sine products), using a more complex set of
$N_{\theta_{[i]}}$ tailored to each specific function allows to reach a
reduction of more than a half of the number of points with even fewer accuracy
loss compared to a uniform mesh. Whether such $N_{\theta_{[i]}}$ set exists for realistic distribution
functions is still unknown.

\subsection{Gyroaverage operator}\label{sec:gyroaverage_res}
Numerical results concerning the verification of the implementation of the gyroaverage operator, described in Section \ref{sec:gyroaverage}, will now be presented. 

A certain family of functions has been considered, whose analytical gyroaverage is known. More precisely, their gyroaverage can be obtained simply by a multiplication of a Bessel function\,\cite{gyroaverage_steiner}. Given the function
\begin{equation}\label{eq:testcase_general}
f(r, \theta) = C_m (zr) \exp{(i m \theta)},
\end{equation}
where $m \ge 0$ is an integer which defines the index of the Bessel function $C_m$ (the symbols $J_m$ and $Y_m$ are used to identify respectively the Bessel functions of the first and the second kind); $z \in \mathbb{C}$ and $(r, \; \theta)$ represent the usual polar coordinates. The gyroaverage of the function described in Eq. \eqref{eq:testcase_general} can be written as in\,\cite{gyroaverage_steiner}:
\begin{equation}
\mathit{J}_\rho(f)(r_0, \theta_0) = J_0(z \rho) C_m(z r_0) \exp{(i m \theta_0)},
\end{equation}
where $\rho$ is the gyroradius, while ($r_0, \theta_0$) define a specific point in the polar mesh.

We have to consider also the boundary conditions. In our case it corresponds to set an homogeneous Dirichlet condition on $r_\text{max}$. We can now list the family of functions we used as test cases. It comes directly from the definition \eqref{eq:testcase_general}, and it's written as:
\begin{equation}\label{eq:testcase1}
f_1(r, \theta) = J_m \Bigl( r \frac{j_{m,\ell}}{r_\text{max}} \Bigr) \exp{(i m \theta)}, \end{equation}
where $j_{m,\ell}$ is the $\ell$-th zero of $J_m$. The function in Eq.~\ref{eq:testcase1} is defined on a disk $[0, r_\text{max}] \times [0,2\pi]$ and verifies the Dirichlet boundary condition for which $f_1(r_\text{max}, \theta) = 0, \quad 0 \le \theta \le 2\pi$. The analytical gyroaverage of \eqref{eq:testcase1} evaluated at the point $(r_0, \theta_0)$ is \cite{gyroaverage_steiner}:
\begin{equation}
\mathit{J}_\rho(f_1)(r_0, \theta_0) = J_0 \Bigl(\rho \frac{j_{m,\ell}}{r_\text{max}} \Bigr) f_1(r_0, \theta_0)
\end{equation}

\begin{figure} 
\includegraphics[width=0.5\textwidth,keepaspectratio]{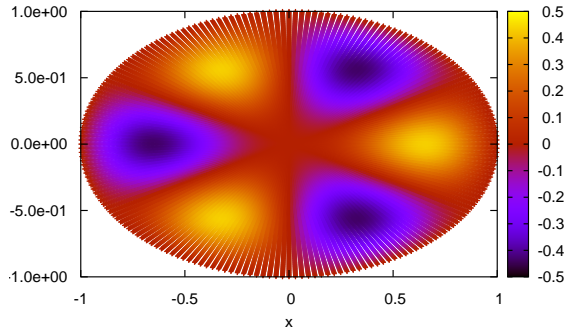} 
\caption{A color map of the test function used for the convergence tests of the gyroaverage operator. The exact expression of the function is 
$f_1(r, \theta) = J_m \Bigl( r \frac{j_{m,\ell}}{r_\text{max}} \Bigr) \exp{(i m \theta)}, $
where the order of the Bessel function considered was $m=3$ and its first zero was used.}
\label{fig:testfunction} \end{figure}

The convergence study results will be shown for the first class of functions presented, described by Eq. \eqref{eq:testcase1}: the order of the Bessel function has been chosen equal to $3$, and the first zero has been considered in the argument. A plot of this function can be seen in Figure~\ref{fig:testfunction}. The uniform and non uniform grid cases have been addressed, and the convergence tests have been performed in both the $r$ and $\theta$ directions as well as in the degree of the underlying Lagrange interpolation. Similar tests have been repeated changing the parameters of the test function, namely the order of the Bessel function and the particular zero of the Bessel function chosen: not all of them are shown here, as the results are very close to the ones presented in this section.

\subsubsection{Convergence tests}

\begin{figure}[th] 
  \begin{subfigure}[b]{0.45\textwidth} 
\hspace*{-.3cm}
\includegraphics[trim= .14cm 0cm .36cm 0cm,clip,width=1.1\textwidth,keepaspectratio]{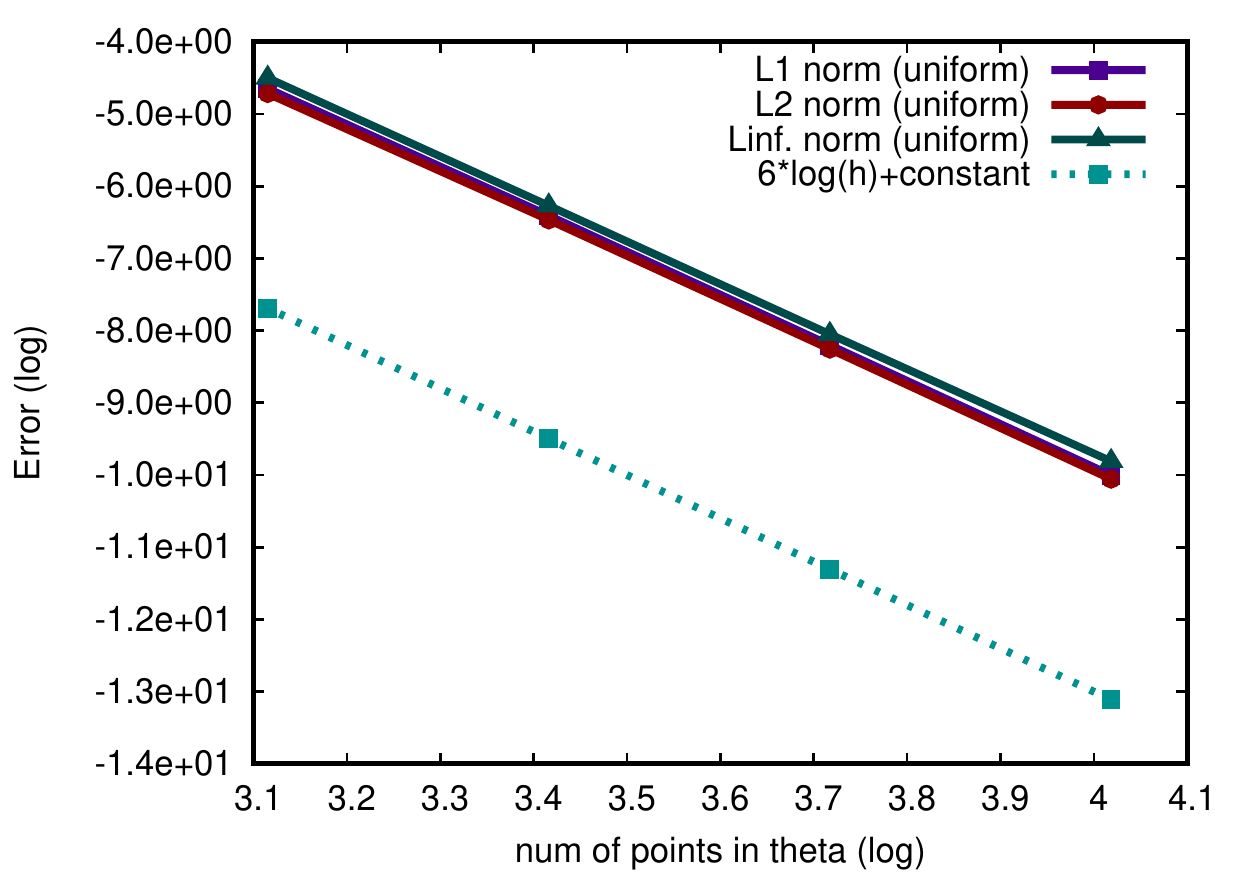}     
\caption{uniform grid} 
\label{fig:gyro_convtest_theta_uniform} \end{subfigure} \hspace{12mm} \begin{subfigure}[b]{0.45\textwidth} 
\hspace*{-.3cm}
\includegraphics[trim= .14cm 0cm .36cm 0cm,clip,width=1.1\textwidth,keepaspectratio]{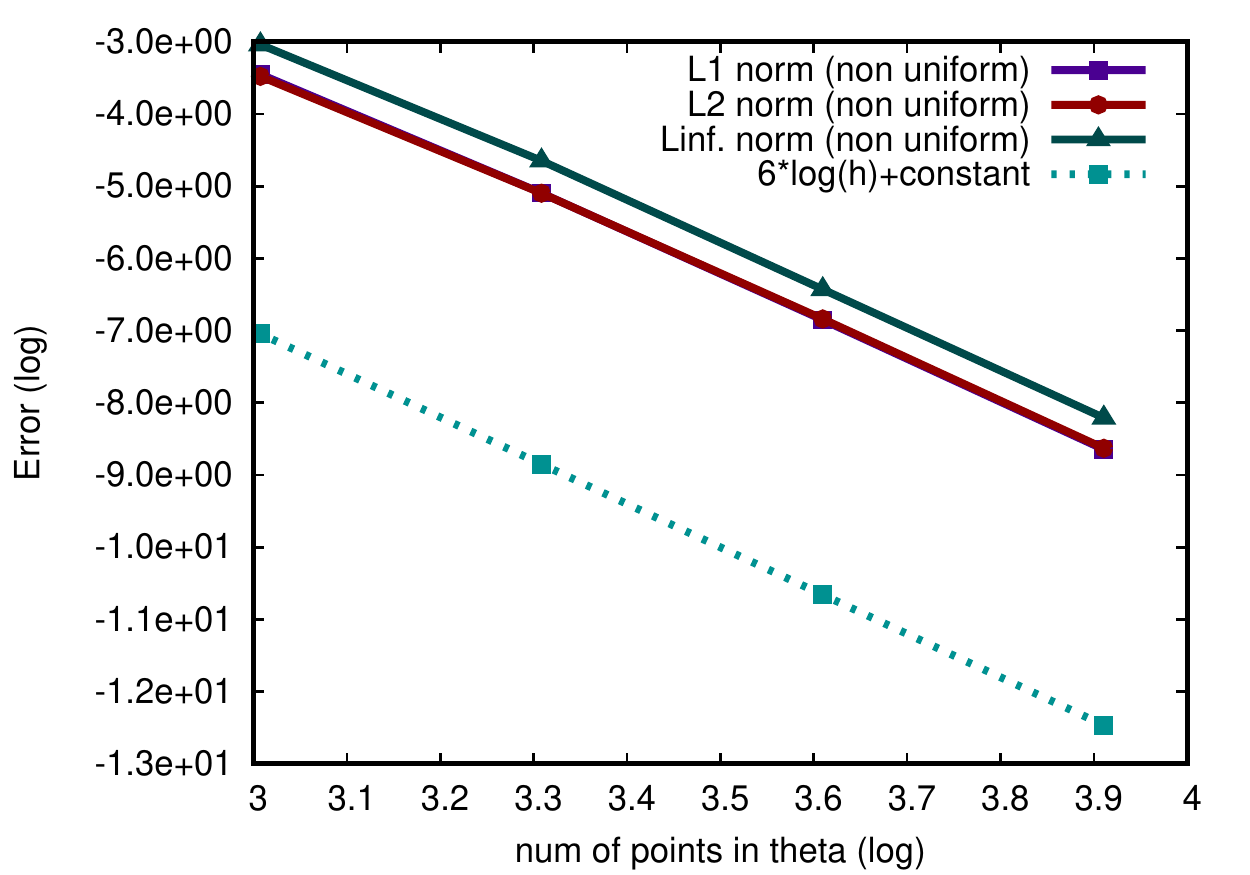}
\caption{non uniform grid}  
\label{fig:gyro_convtest_theta_nonuniform} \end{subfigure} 
\caption{Convergence test for the gyroaverage operator in the $\theta$ direction for the uniform and non uniform grids}
\label{fig:gyro_convtest_theta} \end{figure}

Figure~\ref{fig:gyro_convtest_theta} presents the tests for the convergence in the $\theta$ direction: in particular, the logarithm of the $L^2$ norm, $L^1$ norm and $L^\infty$ norm are shown with respect to the number of points in the $\theta$ direction in a logarithmic scale. 
The degree of the Lagrange interpolation was fixed to be equal to $5$, while the number of points in the $r$ direction and on the gyroaverage circle was respectively equal to $N_r = 256$ and $N_{gp} = 128$, in order to avoid spurious errors related to these parameters; the gyroradius was set equal to $\rho = 0.1$. For the uniform grid (described in Section~\ref{sec:uniform}), the number of points in the $\theta$ direction was respectively $32$, $64$, $128$ and $256$. For the non uniform case however the grid (described in section \ref{sec:non_uniform}) has been built respectively as $(10:\underline{4},30:\underline{8},50:\underline{16},166:\underline{32})$, %
$(10:\underline{8},30:\underline{16},50:\underline{32},166:\underline{64})$, $(10:\underline{16},30:\underline{32},50:\underline{64},166:\underline{128})$, $(10:\underline{32},30:\underline{64},50:\underline{128},166:\underline{256})$, with the same reading convention as described in Section~\ref{sec:interp_res}. This means, for example, $4$ points in the $\theta$ direction have been used for the first $10$ radial positions, $8$ points in the $\theta$ direction have been used for the subsequent $30$ radial positions, and so on. For each of these cases, comparing the uniform and non uniform case, we were thus able to reduce the number of points in the $\theta$ direction by $22\%$.

For both the uniform and non uniform polar meshes we obtain satisfactory convergence results, in accordance with the theoretical expectations given by the blue dashed curve in the following figures. The theoretical slope is in fact given by $n \cdot log(h) + c$, where $h$ is the mesh spacing, $c$ an arbitrary constant and the multiplicative factor $n$ in front of the logarithm is given by $p+1$, with $p$ being the degree of the Lagrange polynomials used for the interpolation.

\begin{figure}[th] 
\begin{subfigure}[b]{0.45\textwidth} 
\hspace*{-.3cm}
\includegraphics[trim= .14cm 0cm .36cm 0cm,clip,width=1.1\textwidth,keepaspectratio]{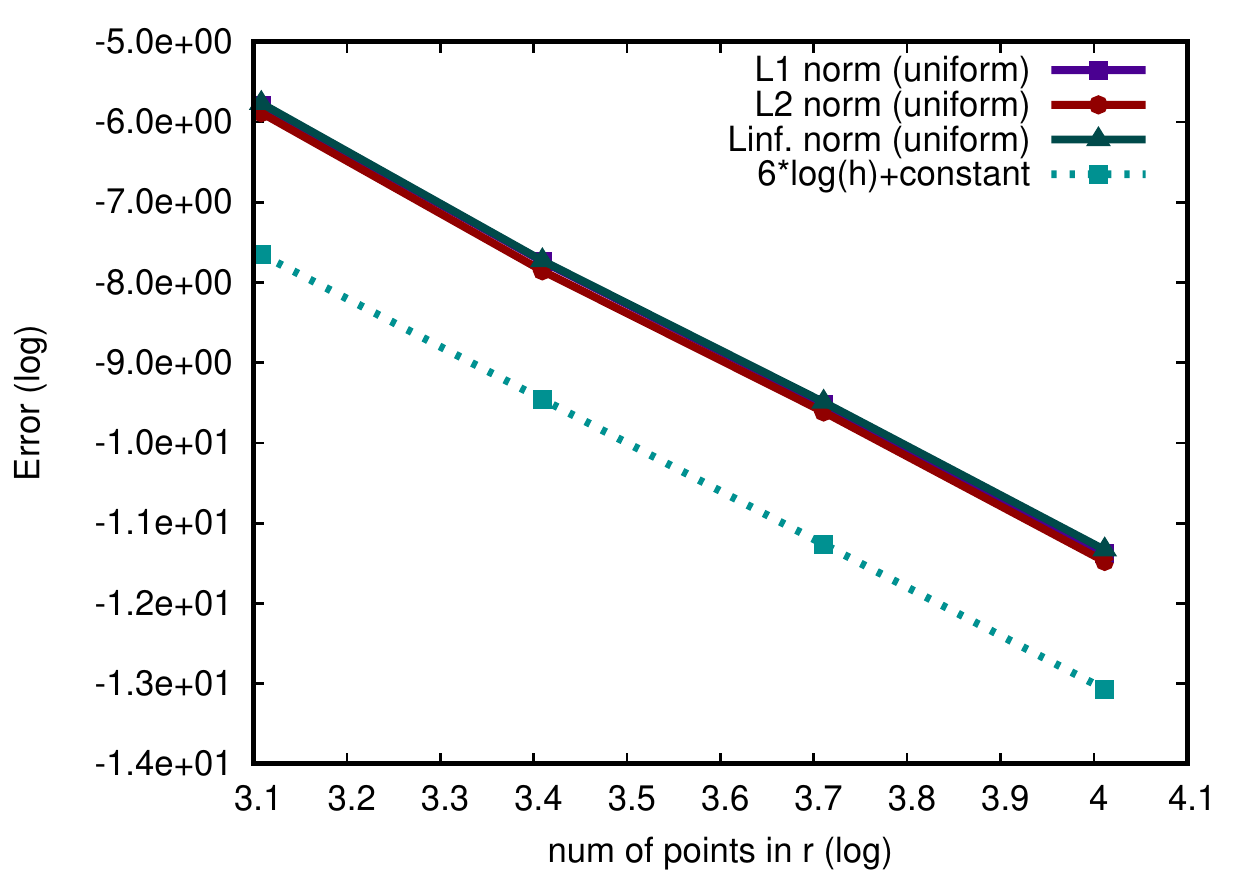}     
\caption{in the $r$ direction} \label{fig:gyro_convtest_r} 
\end{subfigure} \hspace{12mm} \begin{subfigure}[b]{0.45\textwidth} 
\hspace*{-.3cm}
\includegraphics[trim= .14cm 0cm .36cm 0cm,clip,width=1.1\textwidth,keepaspectratio]{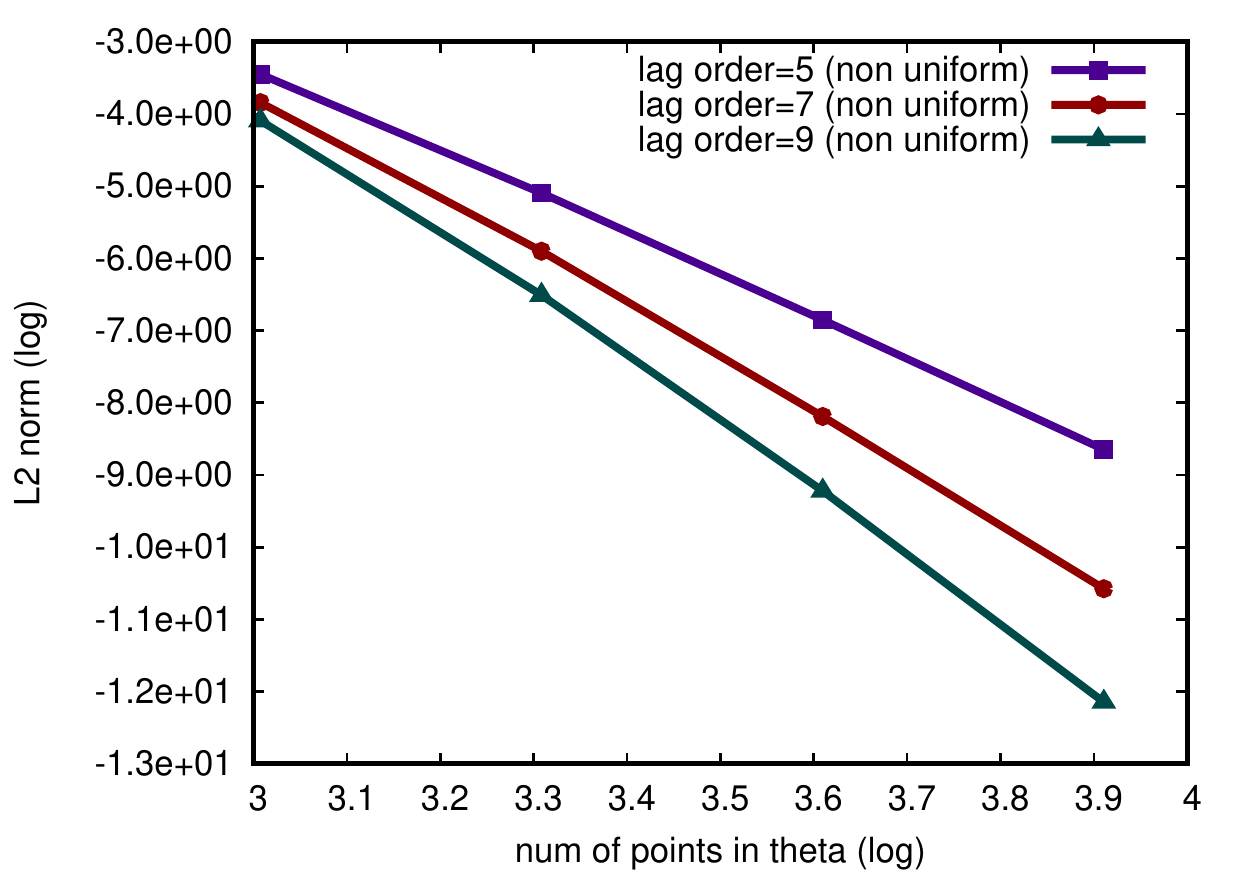}
\caption{on the degree of the Lagrange polynomial}
\label{fig:gyro_convtest_theta_degree} \end{subfigure}
\caption{Convergence test of the gyroaverage operator in the $r$ direction and on the degree of the Lagrange polynomial}
\label{fig:gyro_convtest_r_degree}  \end{figure}

Focusing on the sole logarithm of the $L_2$ norm, the same convergence test described for the Figure \ref{fig:gyro_convtest_theta_nonuniform} has been repeated for different values of the degree of the Lagrange interpolation polynomial. All the other parameters, namely $N_r$, $N_\theta$, $\rho$ and $N_{gp}$, have been kept as in the setting described for the convergence tests in the non uniform $\theta$-direction. The results of this scan is shown in Figure \ref{fig:gyro_convtest_theta_degree}, where the degree of the interpolation has been changed in the range of $(5,7,9)$. 

Figure \ref{fig:gyro_convtest_r} presents the last convergence test which has been performed, namely for the $r$ direction. 
The logarithm of the $L_2$ norm, $L_1$ norm and $L_\infty$ norm are plotted with respect to an increasing number of points in the $r$ direction in a logarithmic scale. Among the parameters kept fixed during the convergence scan in $r$, the degree of the Lagrange polynomial was set equal to~$5$, the number of points uniformly distributed in the $\theta$ direction was equal to $N_\theta = 504$, while the number of points on the gyroaverage circle was set equal to $N_{gp} = 128$. The gyroradius was still considered to be equal to $\rho = 0.1$. The number of points in the $\theta$ direction had to be chosen uniformly and large enough, in order to avoid a contribution of the error due to discretization in the $r$ direction.
The expected theoretical slope, \textit{e.g.} $6\,log(h) + c$, due to the degree $5$ of the polynomial interpolation used, has been observed in the numerical results. 

\subsubsection{Uniform and non uniform case comparison}
For a conclusive comparison between the uniform and non uniform case, the error in the $L^2$ norm has been investigated, given the same number of points in the two grids, consequently differently distributed in the space. In particular, the non uniform sequence grid $(10:\underline{32},30:\underline{64},50:\underline{128},166:\underline{256})$ has been compared with a uniform grid of $200$ points in each radial position, the sequence $(10:\underline{16},30:\underline{32},50:\underline{64},166:\underline{128})$ with a uniform grid of $100$ points for each radial position, the sequence $(10:\underline{8},30:\underline{16},50:\underline{32},166:\underline{64})$ with a grid constituted of $50$ points for each radial position, and finally the non uniform sequence grid $(10:\underline{4},30:\underline{8},50:\underline{16},166:\underline{32})$ with a uniform grid with $25$ points for each radial direction. The other parameters have been kept fixed during the test, and a sufficient amount of points in the $r$ direction ($N_r = 256$) and for the gyroaverage discretization ($N_{gp} = 128$) have again been used in order to avoid possible spurious contributions related to these parameters. Among the other parameters, the degree of the Lagrange interpolation was equal to $5$ and the gyroradius $\rho$ fixed to $0.1$.

\begin{figure}[th] 
\begin{subfigure}[b]{0.45\textwidth}
\hspace*{-.3cm}
\includegraphics[trim= .14cm 0cm .36cm 0cm,clip,width=1.\textwidth,keepaspectratio]{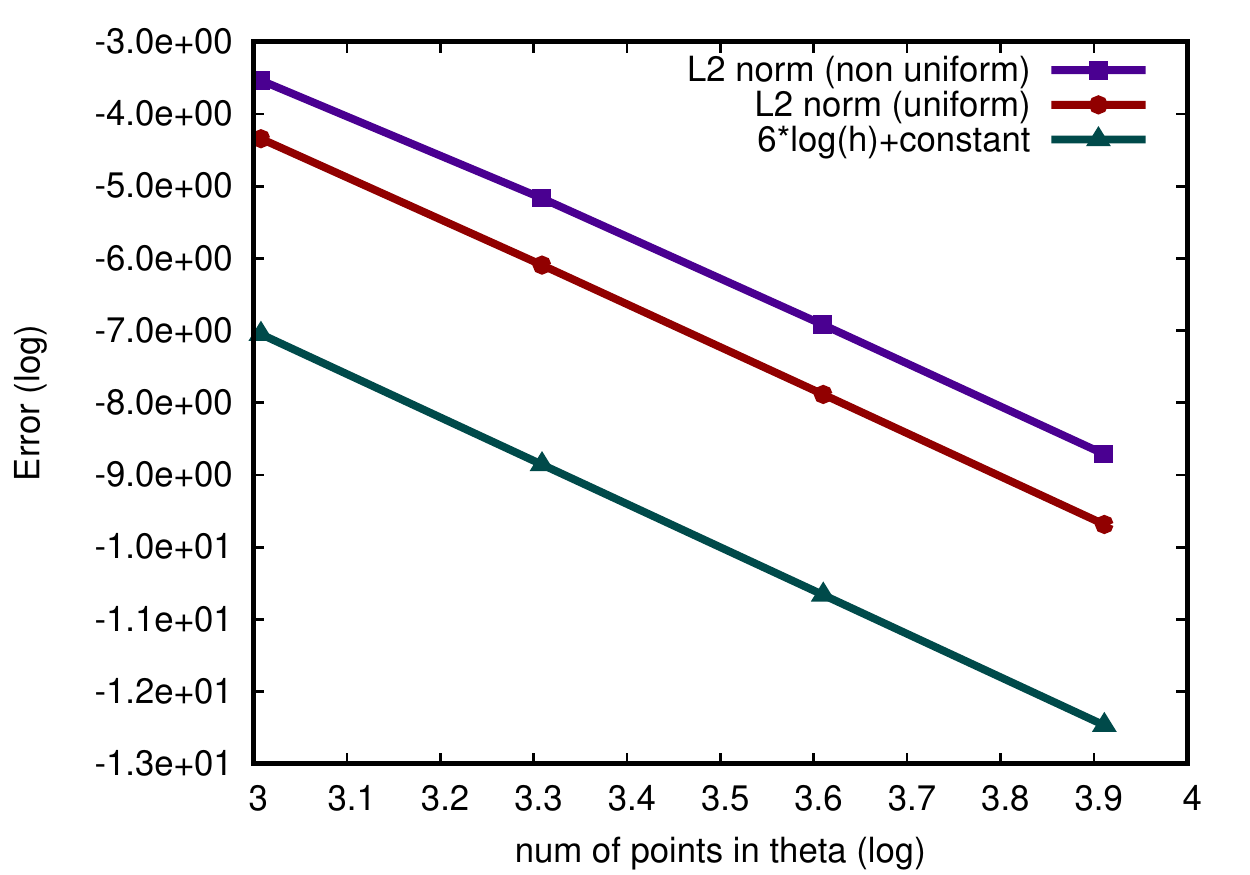}     
\caption{on the domain $r<0.4$} \label{fig:gyro_convtest_theta_extendedregioncomparison} 
\end{subfigure} \hspace{12mm} \begin{subfigure}[b]{0.45\textwidth} \centering  
\hspace*{-.3cm}
\includegraphics[trim= .14cm 0cm .36cm 0cm,clip,width=1.\textwidth,keepaspectratio]{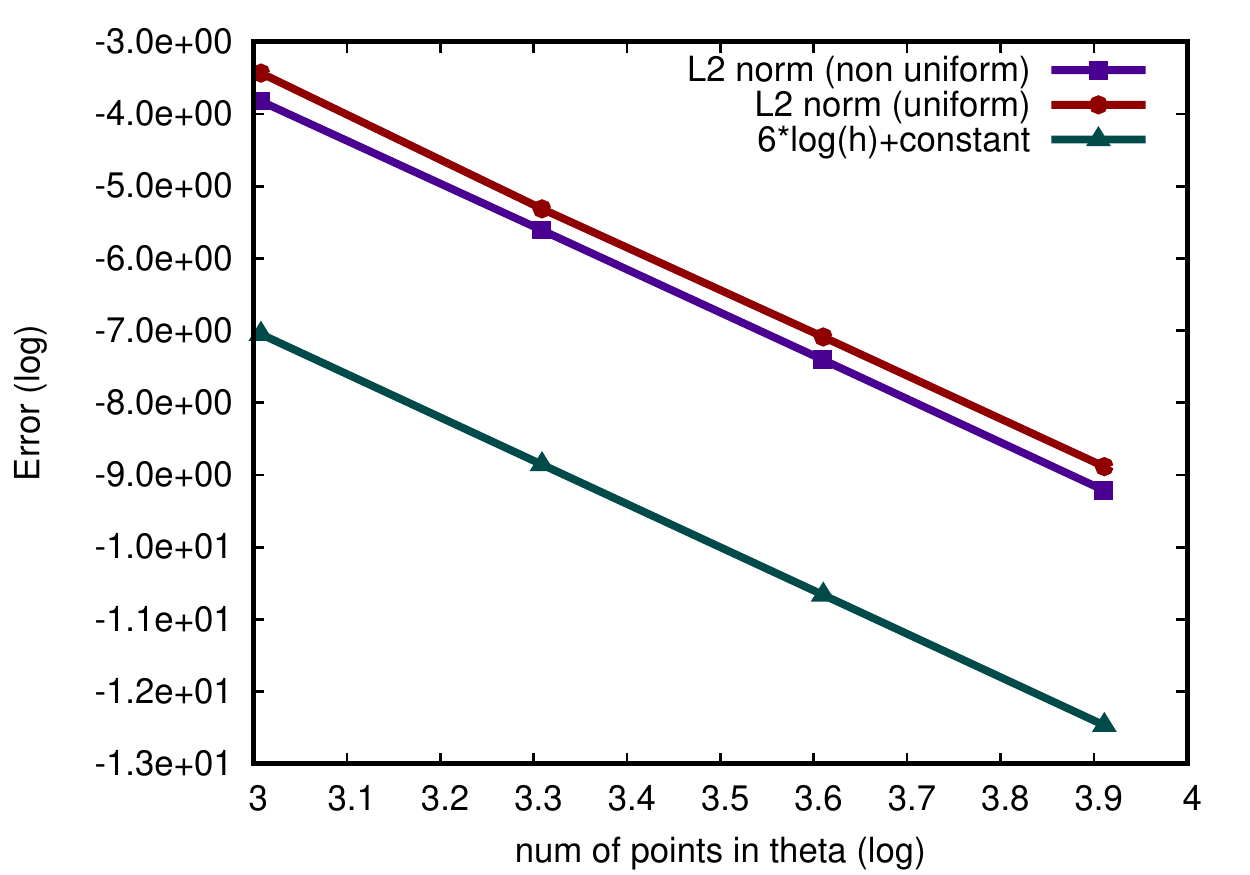}
\caption{on the domain $0.4<r<0.85$}
\label{fig:gyro_convtest_theta_outerregioncomparison} \end{subfigure}
\caption{Error comparison between the uniform and non uniform grid in different regions}
\label{fig:gyro_convtest_theta_comparison}  \end{figure}

The results are shown in Figure \ref{fig:gyro_convtest_theta_comparison}. In Figure \ref{fig:gyro_convtest_theta_extendedregioncomparison} we can see that the error is smaller in the uniform case, as expected, since we are restricting the domain region to radii smaller \mbox{than~$0.4$}. But, if on the other hand we investigate the outer region of the domain ($0.4<r<0.85$), which usually is the one with more need to be accurately resolved due to the interesting physical structures which develop especially here, we can see in Figure \ref{fig:gyro_convtest_theta_outerregioncomparison} that the error is smaller in the non uniform grid case.\\

Given the results of the convergence tests shown in Figures \ref{fig:gyro_convtest_theta} and \ref{fig:gyro_convtest_r_degree}, we can assess that the gyroaverage operator works satisfactorily on the new grid, thus providing a workable implementation. Considering also the last results presented in Figure \ref{fig:gyro_convtest_theta_comparison}, we can conclude that the operator is more accurate on the new non uniform grid in the physically interesting region of the domain, thus proving the usefulness of the method proposed.

\subsection{Advection operator}
\label{sec:advection_res}
Several test cases have been performed for the advection operator. The most
important is the one testing the resilience of structure when advecting through
the center as it is where the mesh is loosen the most. The same meshes as in
\ref{sec:interp_res} are used (\textit{i.e.} 15\% less points on non-uniform
mesh than on uniform mesh). The function studied is the following:

\begin{equation}
    f(r,\theta) = 
    \begin{cases}
      (\cos(1-2\pi\frac{r-r_2}{r_1-r_2}))(1-\cos(2\pi\frac{\theta-\theta_2}{\theta_1-\theta_2})),
      & (r,\theta)\in [r_1,r_2]\times [\theta_1,\theta_2];\\
      0, & \text{otherwise.} \\
    \end{cases}
  \label{eq:advec}
\end{equation}

where $r_{min}\le r_1<r_2 \le r_{max}$ and $0\le \theta_1<\theta_2 \le 2\pi$.

\begin{figure}[!h]
  \centering
  \begin{subfigure}[t]{0.45\textwidth}
    \centering
    \includegraphics[width=1\textwidth]{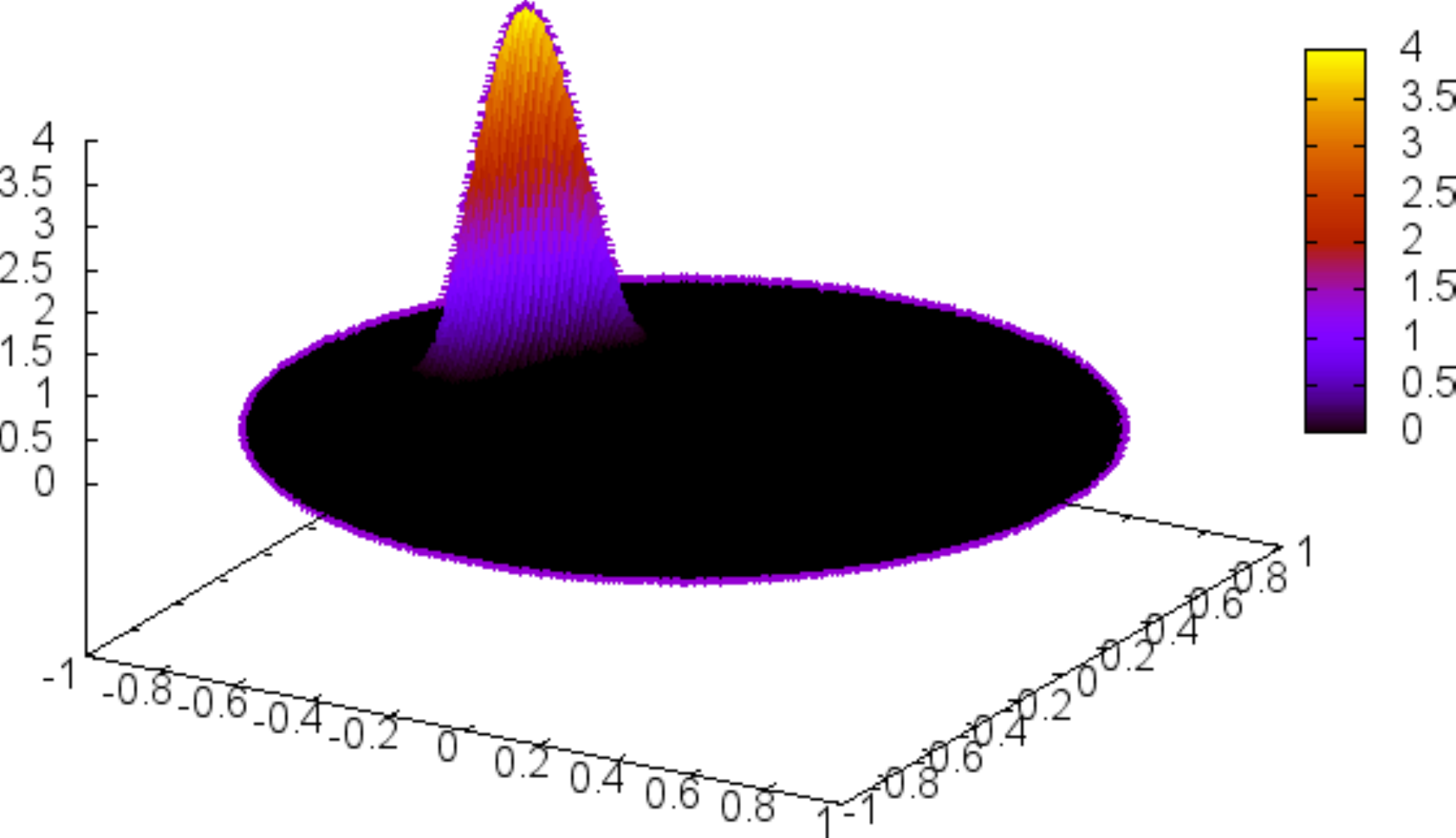}
    \caption{Function at time step 1.}
  \end{subfigure}
  \qquad
  \begin{subfigure}[t]{0.45\textwidth}
    \centering
    \includegraphics[width=1\textwidth]{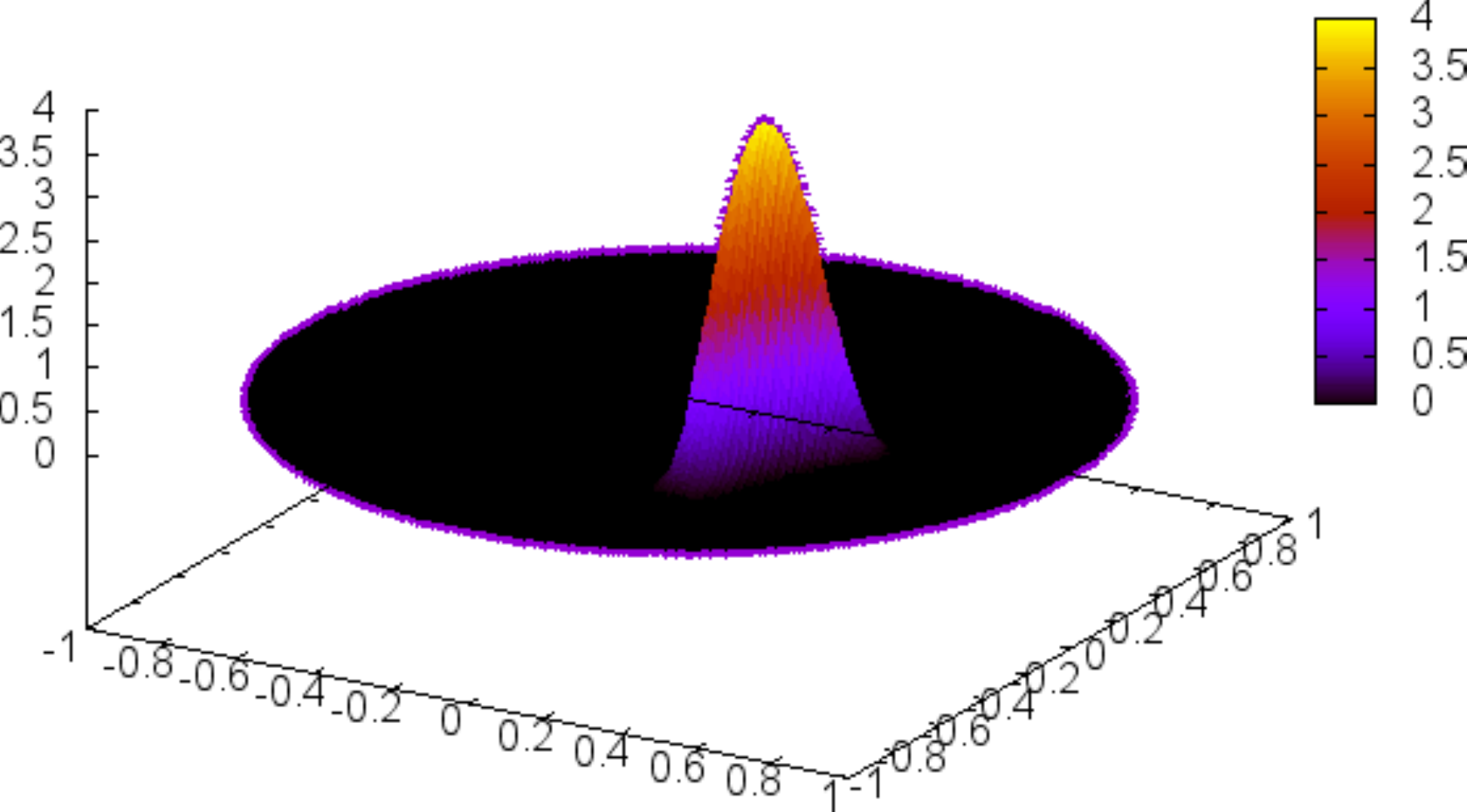}
    \caption{Function at time step 20.}
  \end{subfigure}
  \qquad
  \begin{subfigure}[t]{0.45\textwidth}
    \centering
    \includegraphics[width=1\textwidth]{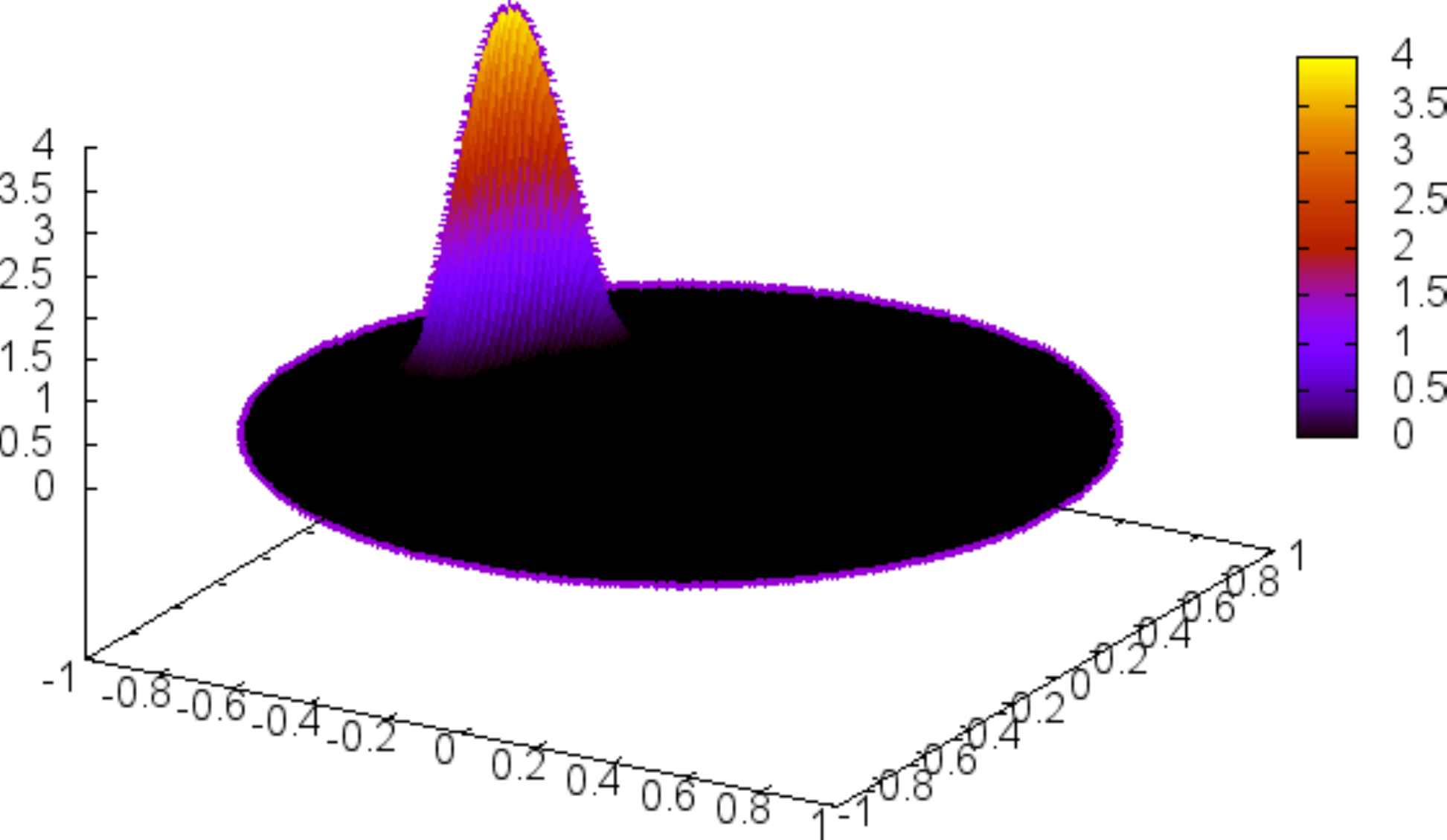}
    \caption{Function at time step 40.}
  \end{subfigure}
  \caption{Evolution of the reference function during the advection.}
  \label{fig:advec_ref}
\end{figure}

\begin{figure}[!h]
  \centering
  \begin{subfigure}[t]{0.45\textwidth}
    \centering
    \includegraphics[width=1\textwidth]{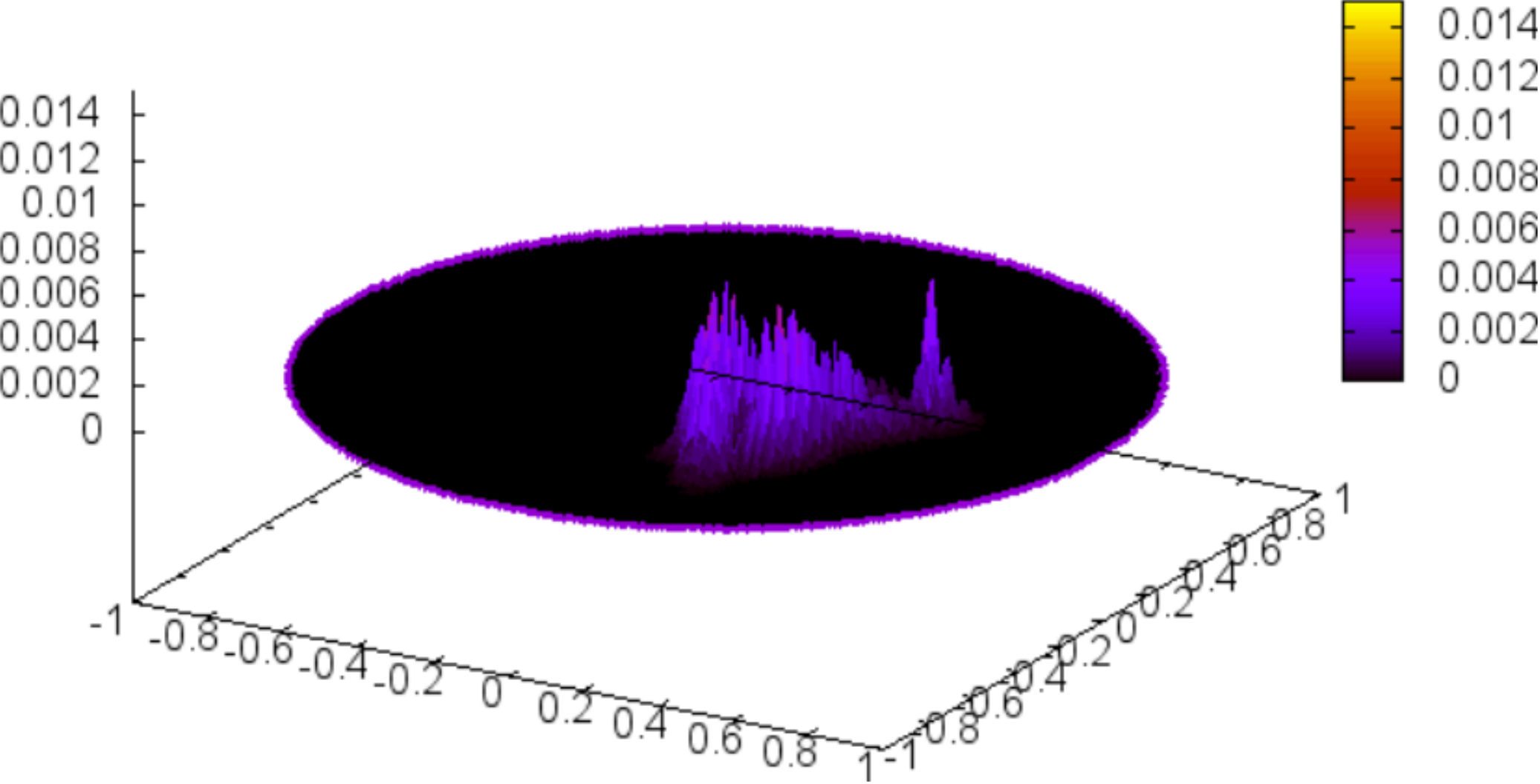}
    \caption{Error at time step 20 on uniform mesh.}
    \label{fig:advec_uni_20}
  \end{subfigure}
  \qquad
  \begin{subfigure}[t]{0.45\textwidth}
    \centering
    \includegraphics[width=1\textwidth]{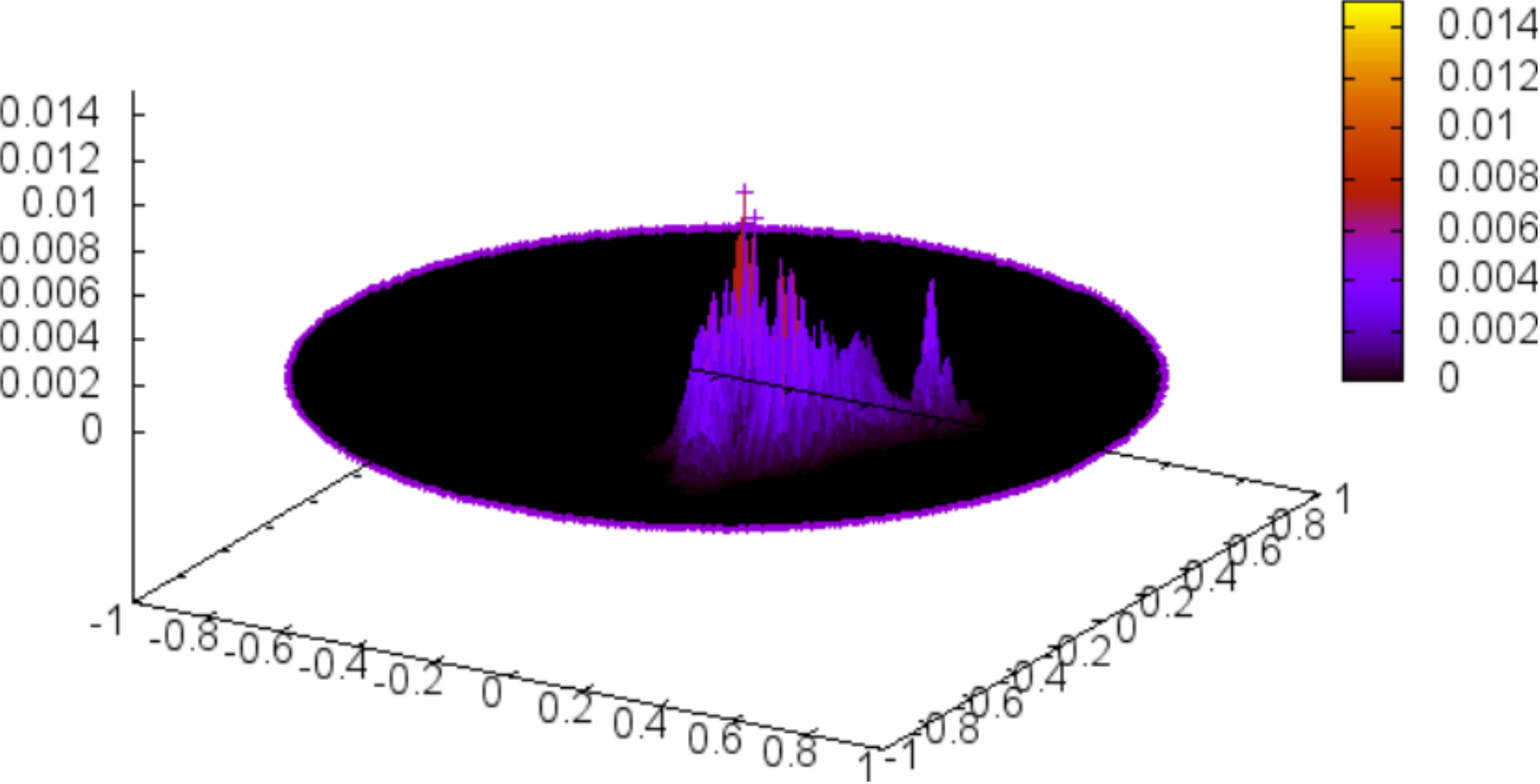}
    \caption{Error at time step 20 on non-uniform mesh.}
    \label{fig:advec_non_uni_20}
  \end{subfigure}
  \qquad
  \begin{subfigure}[t]{0.45\textwidth}
    \centering
    \includegraphics[width=1\textwidth]{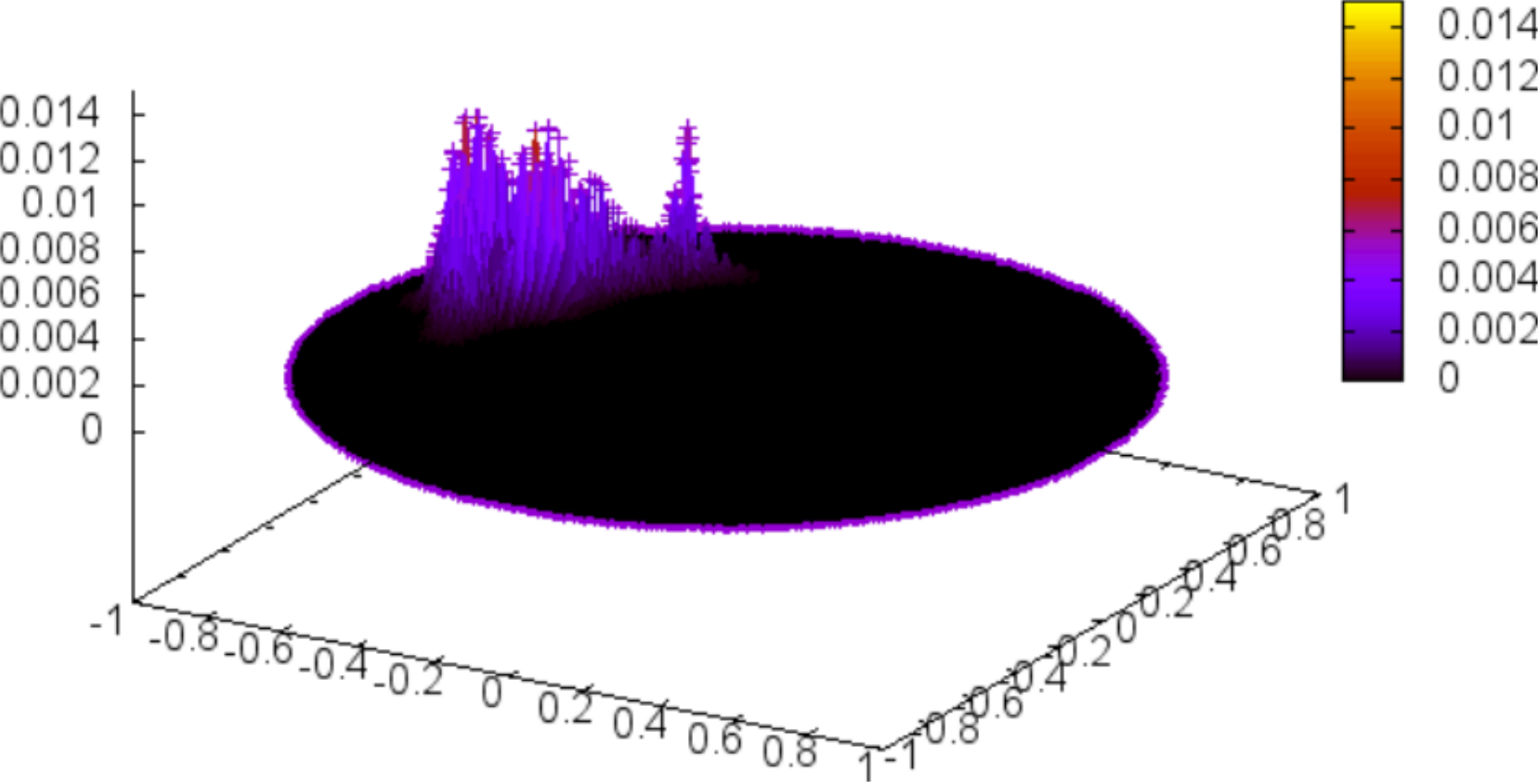}
    \caption{Error at time step 40 on uniform mesh.}
    \label{fig:advec_uni_40}
  \end{subfigure}
  \qquad
  \begin{subfigure}[t]{0.45\textwidth}
    \centering
    \includegraphics[width=1\textwidth]{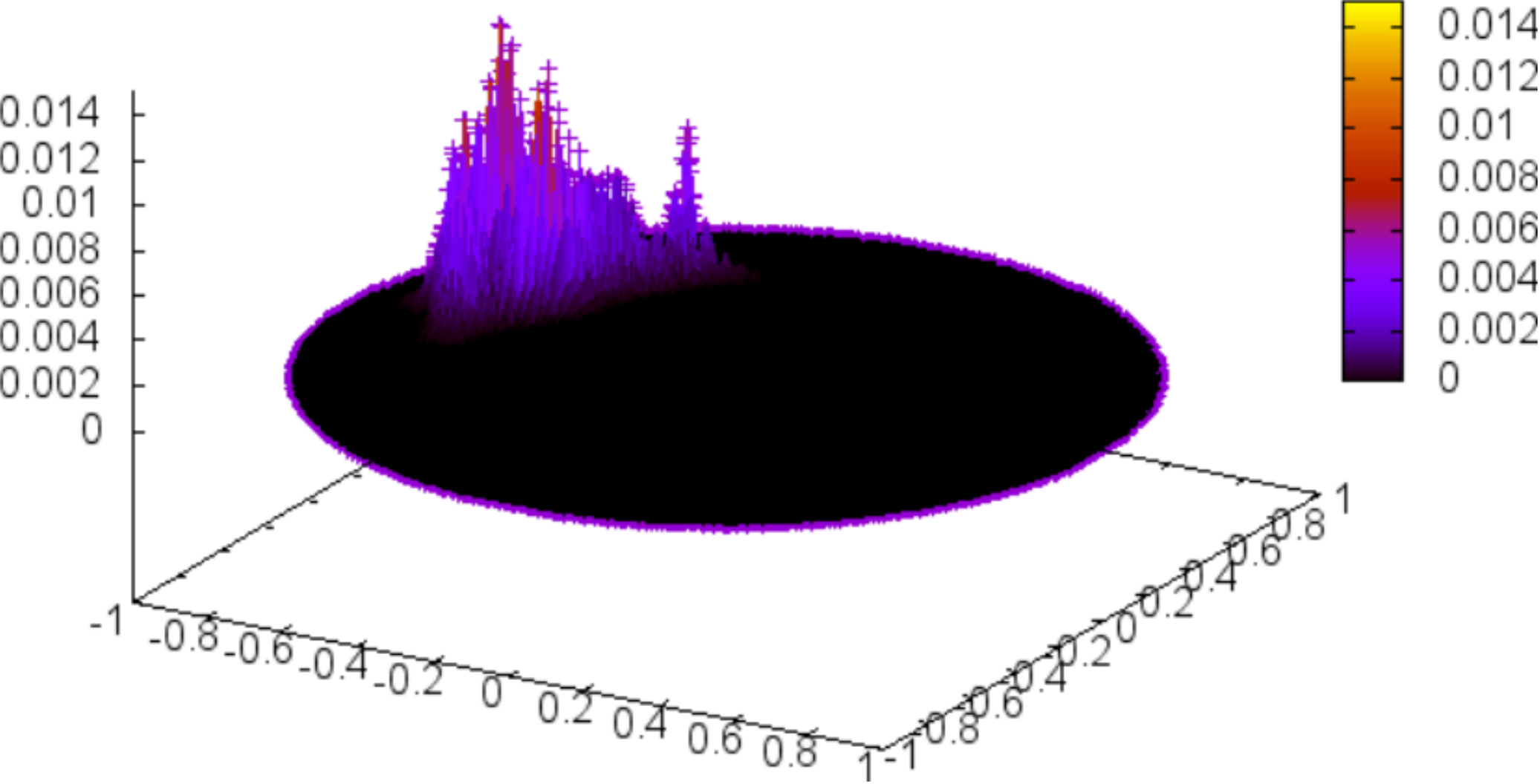}
    \caption{Error at time step 40 on non-uniform mesh.}
    \label{fig:advec_non_uni_40}
  \end{subfigure}
  \caption{Evolution of the error for an advection on uniform and non-uniform mesh.}
  \label{fig:advec}
\end{figure}

This ensures $\mathcal{C}^1$ continuity for Lagrangian interpolation. The
function is then centered on $(r=0.7,\theta=\frac{6\pi}{7})$ ($r_1=0.6$,
$r_2=0.8$, $\theta_1=\frac{5\pi}{7}$, $\theta_2=\pi$), with $r_{max}=1.0$ and
then advected through the center with speed $(v_x=\frac{2}{3},v_y=-\frac{1}{3})$
until $r=0.7$ is reached on the other side. The opposite advection is finally
performed to bring back the structure to its original position. The overall
displacement is performed in 40 time steps. Initial, middle and final snapshots
of the advected function are presented on Figure \ref{fig:advec_ref}. Middle and
final snapshots of the error are presented on Figure \ref{fig:advec_uni_20} and
\ref{fig:advec_uni_40} for the uniform mesh and on Figure
\ref{fig:advec_non_uni_20} and \ref{fig:advec_non_uni_40} for the non-uniform
mesh. These show the difference between the exact value and the interpolated one
for the whole plane at time steps 1, 20 and 40. During the advection, the error
done when interpolating quickly grows when approaching and going through the
center (from almost flat error at time step 1 to \ref{fig:advec_uni_20}). But it
does not evolve on the outer \textit{radii} of the mesh, nor when going back
through the center. It is more pronounced for the non-uniform mesh.

\begin{figure}[!h]
  \centering
  \includegraphics[width=0.5\textwidth]{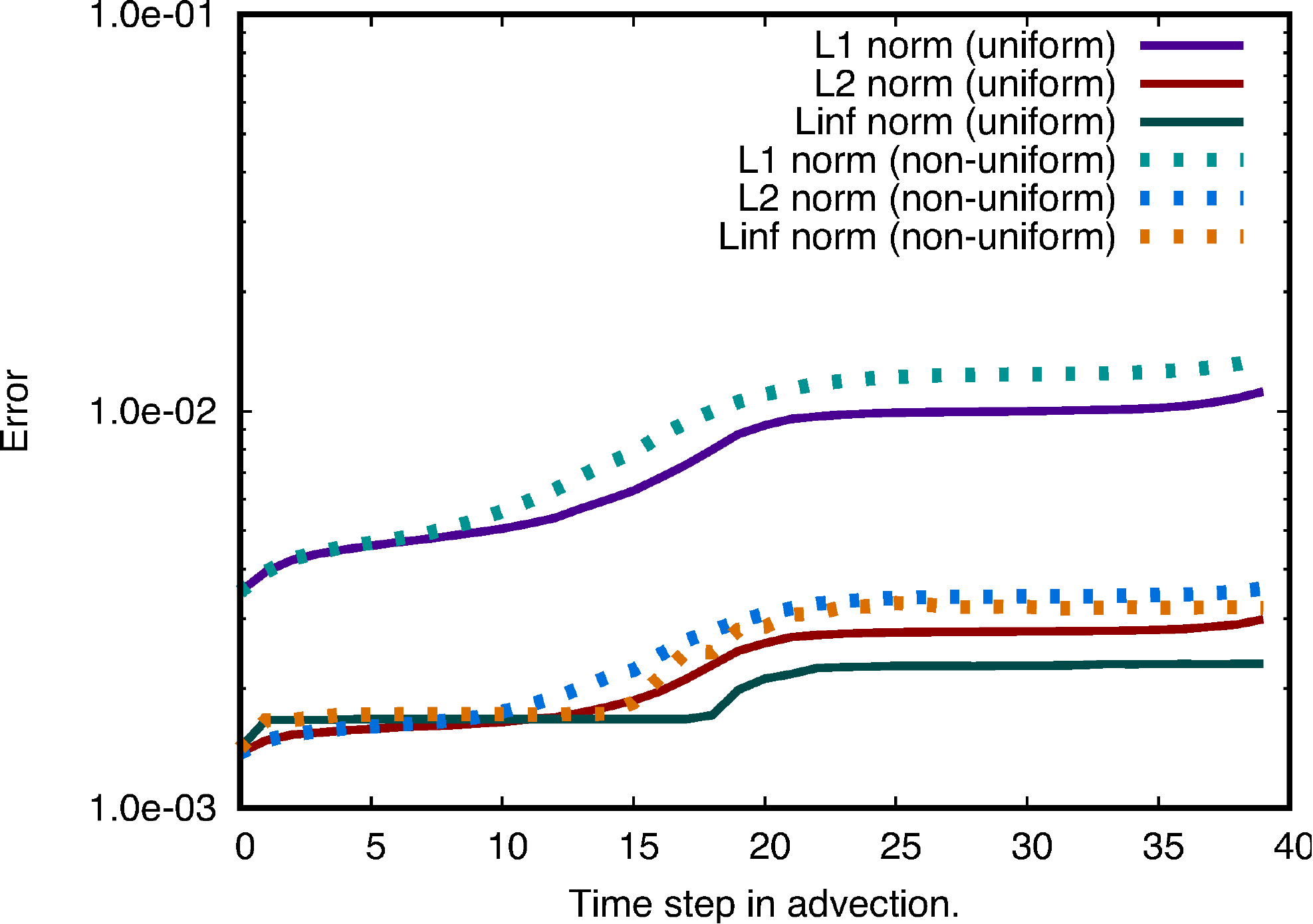}
  \caption{Evolution of mean error when advecting a structure forth and back
    through the center of the plane.}
  \label{fig:advec_graph}
\end{figure}

This is best seen in Figure \ref{fig:advec_graph} which gives the $L_1$ and
$L_2$ norms of the error for the whole plane at each time step for both
meshes. The structure first undergoes accuracy loss when entering the center
near time step 12 for the uniform mesh and near time step 8 for the non-uniform
mesh. This means that one of the $N_{\theta_{[i]}}$ in that region was not high
enough to reach the corresponding accuracy. \textit{Radii} closer to the center did not
have a $N_{\theta_{[i]}}$ high enough either as shows the higher overall error
for the non-uniform mesh. Once the structure has gone once through the coarsely
solved part of the mesh, the accuracy does not undergo any other drastic
reduction anymore. It means we have reached the minimum resolution offered by
the mesh. This result is satisfactory.

Once again, the operator works as expected. The accuracy highly depends on the
choice of the $N_{\theta_{[i]}}$, especially near the center where we wanted the number of points to be scarce. It could be useful to have a tuning tool, which, given the desired
accuracy, the number of \textit{radii} and the typical variation of the function
(size of the smallest structure to solve), would give the \mbox{optimal~$N_{\theta_{[i]}}$}, but we do not have it yet.

\paragraph{Results with large aspect ratio mapping}

\begin{figure}[!h]
  \hspace*{-0.9cm}
  \begin{subfigure}[t]{0.32\textwidth}
    \centering
    \includegraphics[width=1\textwidth]{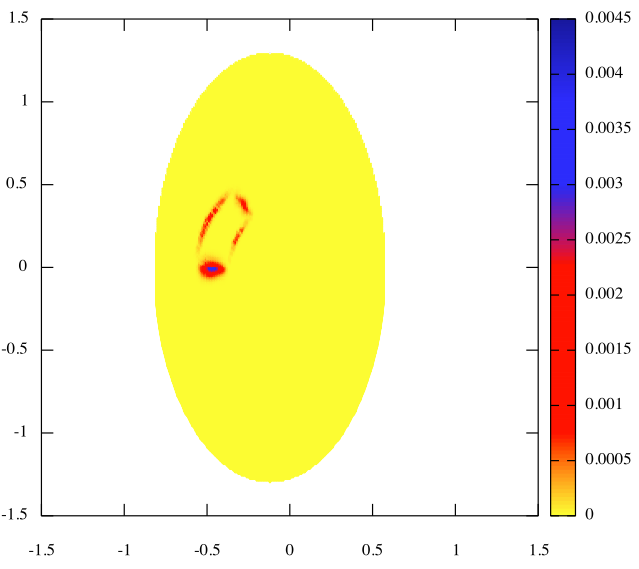}
    \caption{Uniform mesh}
    \label{fig:advec_culham_uni_1}
  \end{subfigure}
  \begin{subfigure}[t]{0.32\textwidth}
    \centering
    \includegraphics[width=1\textwidth]{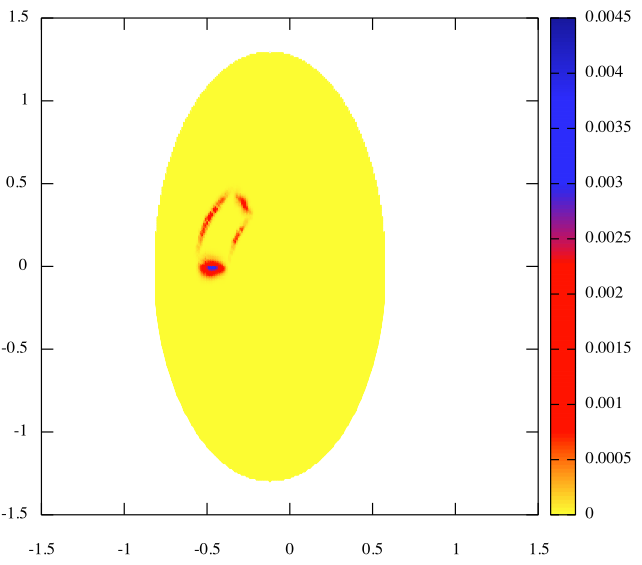}
    \caption{First non uniform mesh}
    \label{fig:advec_culham_non_uni_nb_1}
  \end{subfigure}
  \begin{subfigure}[t]{0.32\textwidth}
    \centering
    \includegraphics[width=1\textwidth]{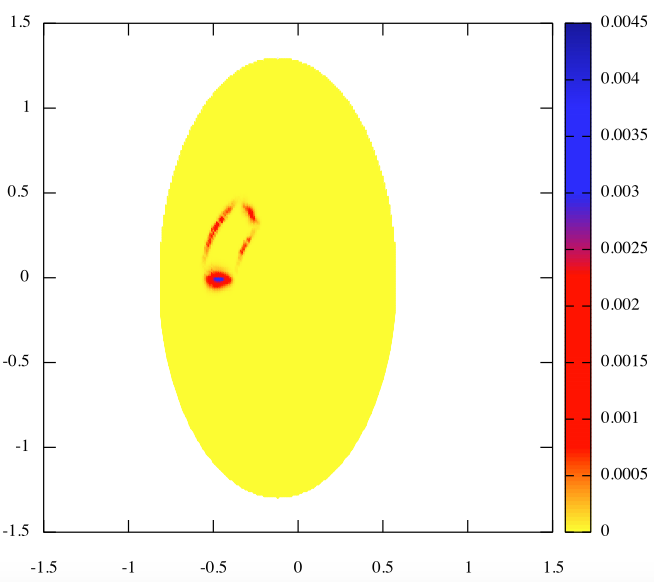}
    \caption{Second non uniform mesh}
    \label{fig:advec_culham_non_uni_cb_1}
  \end{subfigure}
  \caption{Error at time step 1, for an advection on different meshes (large aspect ratio mapping)}
  \label{fig:advec_culham_time1}
\end{figure}

\begin{figure}[!h]
  \hspace*{-0.9cm}
  \begin{subfigure}[t]{0.32\textwidth}
    \centering
    \includegraphics[width=1\textwidth]{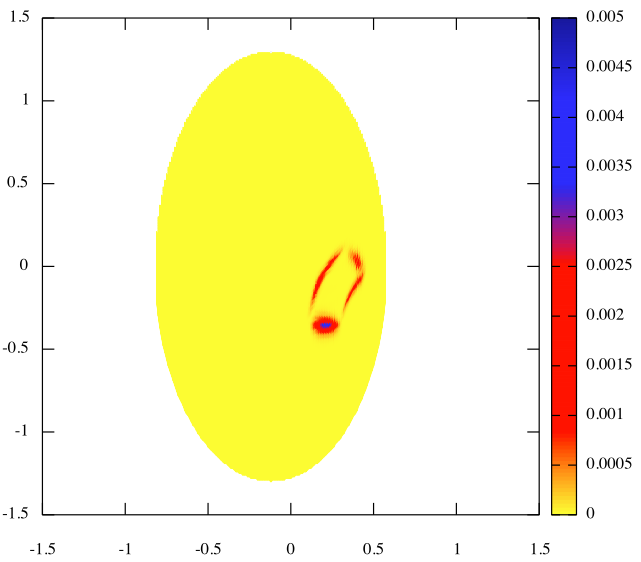}
    \caption{Uniform mesh}
    \label{fig:advec_culham_uni_20}
  \end{subfigure}
  \begin{subfigure}[t]{0.32\textwidth}
    \centering
    \includegraphics[width=1\textwidth]{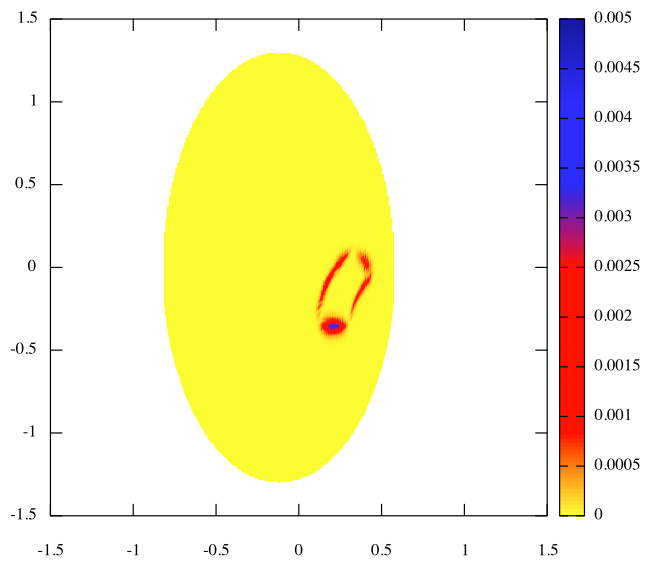}
    \caption{First non uniform mesh}
    \label{fig:advec_culham_non_uni_nb_20}
  \end{subfigure}
  \begin{subfigure}[t]{0.32\textwidth}
    \centering
    \includegraphics[width=1\textwidth]{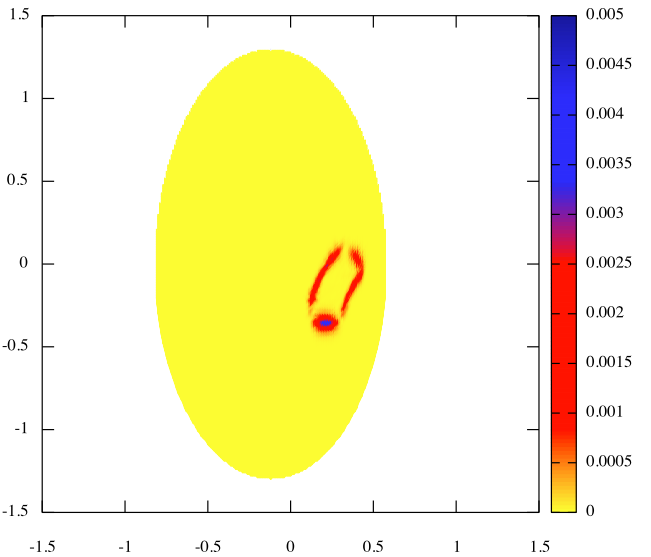}
    \caption{Second non uniform mesh}
    \label{fig:advec_culham_non_uni_cb_20}
  \end{subfigure}
  \caption{Error at time step 20, for an advection on different meshes (large aspect ratio mapping)}
  \label{fig:advec_culham_time20}
\end{figure}

\begin{figure}[!h]
  \hspace*{-0.9cm}
  \begin{subfigure}[t]{0.32\textwidth}
    \centering
    \includegraphics[width=1.\textwidth]{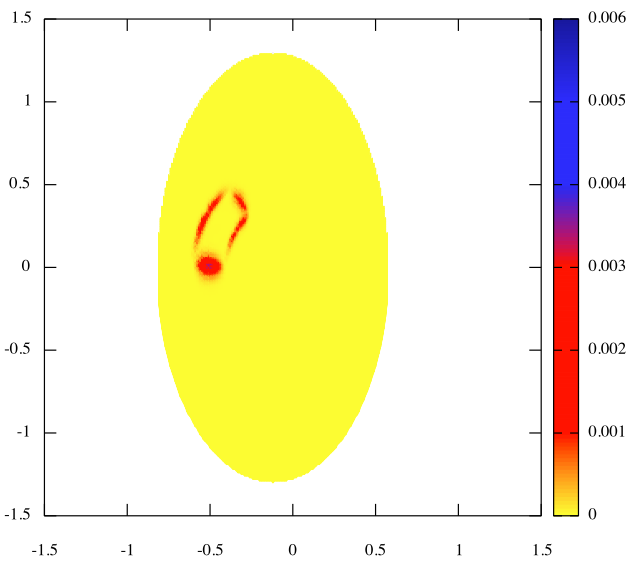}
    \caption{Uniform mesh}
    \label{fig:advec_culham_uni_40}
  \end{subfigure}
  \begin{subfigure}[t]{0.32\textwidth}
    \centering
    \includegraphics[width=1.\textwidth]{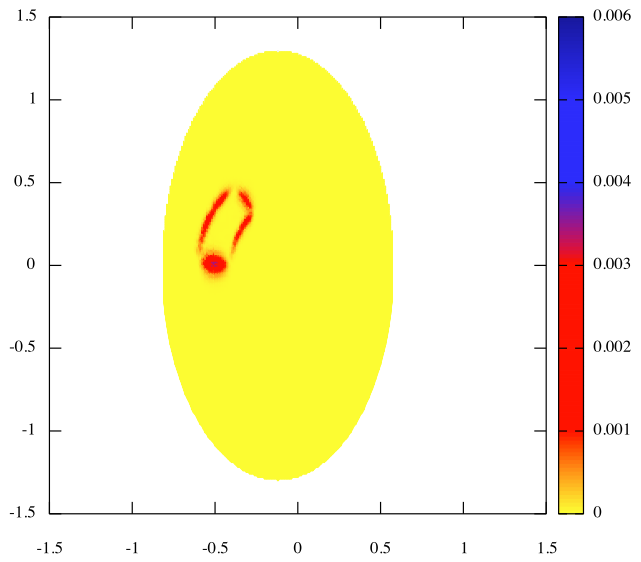}
    \caption{First non uniform mesh}
    \label{fig:advec_culham_non_uni_nb_40}
  \end{subfigure}
  \begin{subfigure}[t]{0.32\textwidth}
    \centering
    \includegraphics[width=1.\textwidth]{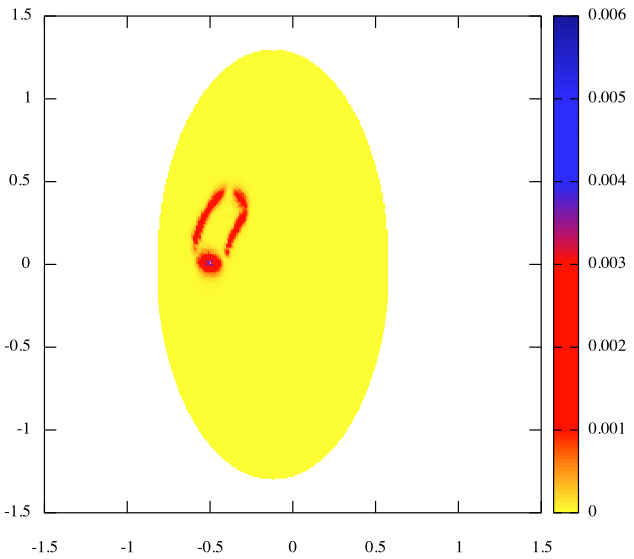}
    \caption{Second non uniform mesh}
    \label{fig:advec_culham_non_uni_cb_40}
  \end{subfigure}
  \caption{Error at time step 40, for an advection on different meshes (large aspect ratio mapping)}
  \label{fig:advec_culham_time40}
\end{figure}

\begin{figure}[!h]
 \begin{subfigure}[t]{0.45\textwidth}
  \includegraphics[width=\textwidth]{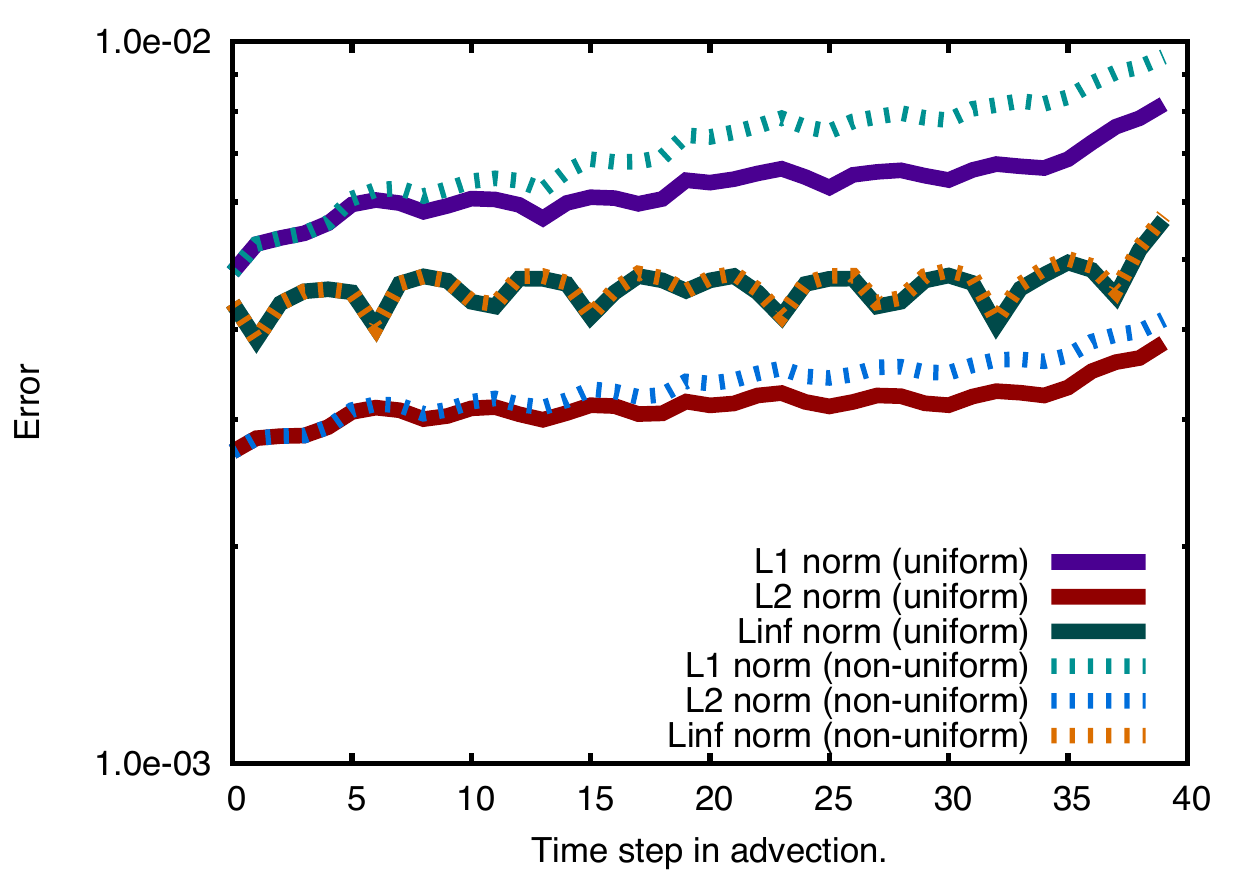}
  \caption{Evolution of mean error when advecting a structure forth and back
    through the center of the plane (large aspect ratio mapping case, with first non-uniform mesh)}
  \label{fig:advec_graph_culham}
 \end{subfigure}
 \qquad
 \begin{subfigure}[t]{0.45\textwidth}
  \includegraphics[width=\textwidth]{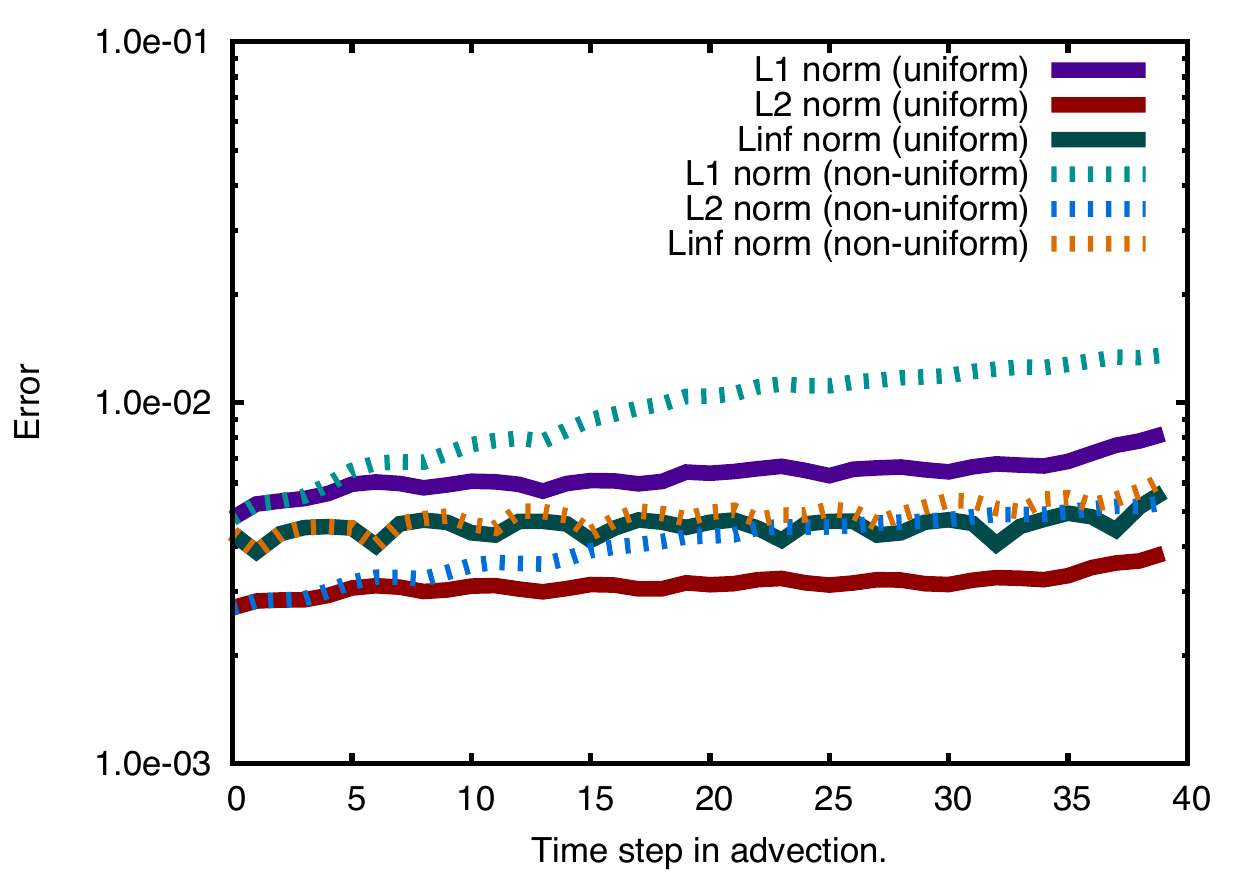}
  \caption{Evolution of mean error when advecting a structure forth and back
    through the center of the plane (large aspect ratio mapping, with second non-uniform mesh)}
  \label{fig:advec_graph_culham2}
 \end{subfigure}
 \caption{Evolution of mean error for the advection}
\end{figure}

Results on the large aspect ratio mapping, for the advection are reported on Figures \ref{fig:advec_culham_time1},\ \ref{fig:advec_culham_time20},\ \ref{fig:advec_culham_time40} and  
\ref{fig:advec_graph_culham},\ \ref{fig:advec_graph_culham2}. 


The results are very similar to the previous polar case. We had to change a little the test case, so that the
solution does not go outside the domain. The initial condition is now in polar coordinates:
$$
f(r,\theta) = \left(1 - \cos(2\pi\frac{r-0.4}{0.2})\right)\left(1-\cos(2\pi\frac{\theta-5\pi/7}{2\pi/7})\right),\ 0.4 \le r\le 0.6,\ \frac{5\pi}{7}\le \theta \le \frac{7\pi}{7} 
$$
and $f(r,\theta)=0$ elsewhere. We do $40$ time steps and use here $v_x=\frac{5}{7}\frac{4}{40},\ v_y=-\frac{5}{7}\frac{2}{40}$ and time step $dt=\frac{1}{2}$ for the $20$ first time steps
and then $v_x=-\frac{5}{7}\frac{4}{40},\ v_y=\frac{5}{7}\frac{2}{40}$ and time step $dt=\frac{1}{2}$ for the $20$ last time steps.

\noindent The $L^1$ and $L^2$ norms are computed as in the polar case (we do not include the Jacobian corresponding to the new mapping here).

\noindent We have used here $N_r=256$, $N_\theta=256$ for the uniform grid and the sequence $(2:\underline{32},8:\underline{64},64:\underline{128},182:\underline{256})$
for the first non uniform grid and the sequence $(10:\underline{32},30:\underline{64},50:\underline{128},166:\underline{256})$ for the second non uniform mesh. 


\subsection{Numerical results for Poisson solver}
\label{sec:poisson_res}
In the following, the previous Poisson Equation \eqref{eq:poiss_polar} is
 solved for the right hand side $R(r,\theta)=2*\exp(r(\cos(\theta)+\sin(\theta)))$ on the polar domain $\Omega=\{(r,\theta):0<r<1 \text{ and } 0\le \theta \le 2\pi \}$. The linear sparse matrix system \eqref{eq:A_matrix_nonunif}-\eqref{eq:b_nonunif} is solved by using the Intel PARDISO solver\footnote{\url{http://pardiso-project.org/}} which is available through the INTEL MKL library. The number of threads (\texttt{MKL$\_$NUM$\_$THREADS}) has been chosen equal to 8. 
All the following simulations  have been performed on a SandyBridge machine (Intel E5-2670 v1, 2.60GHz, 8-cores per socket, 2 sockets per node).  
Numerical results have been compared with analytic results $f(r,\theta)=\exp(r(\cos(\theta)+\sin(\theta))) = 0.5*R(r,\theta)$. 
 
\subsubsection{Reference case with uniform mesh}
\label{sec:poisson_res_ref}
\begin{figure}[h!]
  \begin{minipage}{0.45\linewidth}
    \centering
    \includegraphics[width=1.2\linewidth]{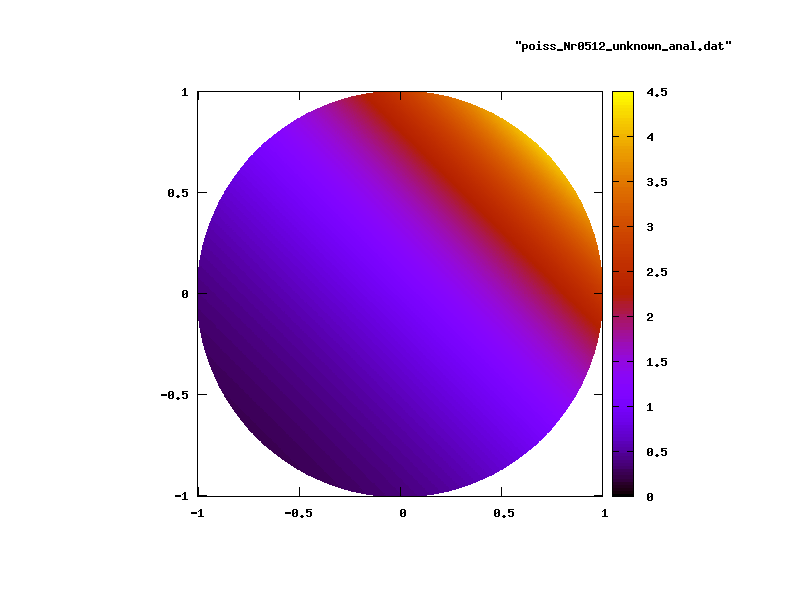}
    \caption{Analytic solution $f(r,\theta)=\exp(r(\cos(\theta)+\sin(\theta)))$ for the case $N=512$ and $N_\theta=1024$.}
    \label{fig:anal_poiss_uniform_Nr512}
  \end{minipage}
  \qquad
  \begin{minipage}{0.45\linewidth}
    \centering
    \includegraphics[width=1.2\linewidth]{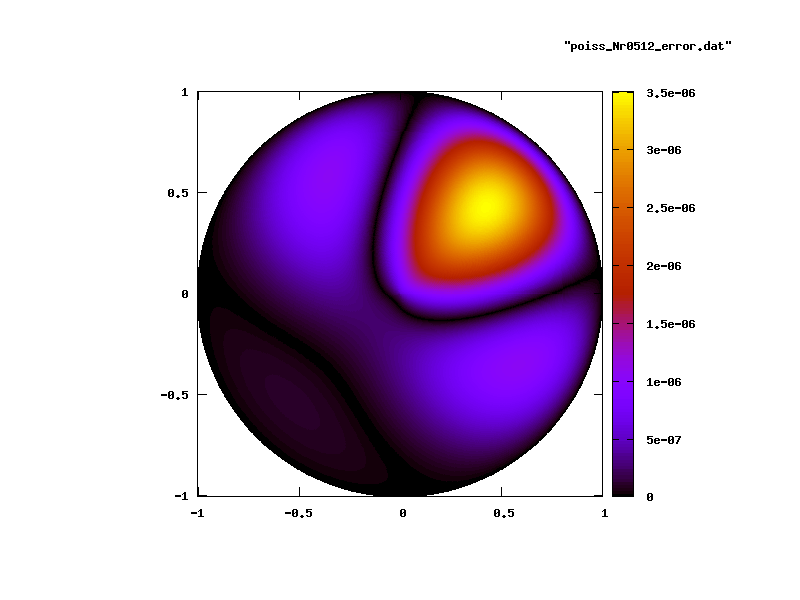}
    \caption{Relative error between numerical results and analytic solution for the case $N=512$ and $N_\theta=1024$.}
    \label{fig:error_poiss_uniform_Nr512}
  \end{minipage}
\end{figure}

First tests have been performed on a uniform mesh where the number of poloidal points $\Ntheta$ has been fixed equal to $\Ntheta=2 N$  ($N$ being the radial cell number). The analytic solution is plotted for the case $N=512$ on Figure \ref{fig:anal_poiss_uniform_Nr512} and the corresponding relative error for the numerical solution is shown in Figure \ref{fig:error_poiss_uniform_Nr512}. The maximum relative error depending on the mesh discretization is summarized in Table \ref{tab:mesh_uniform} for $N$ varying from $8$ to $2048$. Table \ref{tab:mesh_uniform} also shows that the numerical scheme is as expected of second order. These results act as reference results for the next tests on non-uniform meshes. Let us notice that the results on a uniform mesh are the same than the one obtained by Lai \cite{lai2001note}. As said in section \ref{sec:poisson}, the matrix system is constructed in the opposite way compared to the one proposed by Lai in order to be able to extend it to meshes with non-uniform number of poloidal points. Both matrix systems have been also compared in terms of CPU time and are equivalent (results for our matrix system are recorded in Table \ref{tab:mesh_uniform}).

\begin{table}[!h]
  \begin{tabular}[h]{c|c|c|c|c}
    \multicolumn{3}{c}{}        & \multicolumn{2}{|c}{CPU time} \\
    $N$ & Maximum Error & Order & precompute & solve \\
    \hline
    & & & \vspace{-0.2cm}\\
    $4$    &  $5.6814\;10^{-2}$ &         & $0.006607 \,s$  & $0.001120 \,s$ \\
    $8$    &  $1.3950\;10^{-2}$ & $2.026$ & $0.001959 \,s$  & $0.000124 \,s$ \\
    $16$   &  $3.5244\;10^{-3}$ & $1.985$ & $0.005970 \,s$  & $0.000990 \,s$ \\
    $32$   &  $8.8807\;10^{-4}$ & $1.988$ & $0.010281 \,s$  & $0.000315 \,s$ \\
    $64$   &  $2.2261\;10^{-4}$ & $1.996$ & $0.037393 \,s$  & $0.000451 \,s$ \\
    $128$  &  $5.5755\;10^{-5}$ & $1.997$ & $0.167770 \,s$  & $0.001448 \,s$ \\ 
    $256$  &  $1.3950\;10^{-5}$ & $1.999$ & $0.749632 \,s$  & $0.006083 \,s$ \\ 
    $512$  &  $3.4890\;10^{-6}$ & $1.999$ & $3.413063 \,s$  & $0.024860 \,s$ \\ 
    $1024$ &  $8.7244\;10^{-7}$ & $1.999$ & $16.345875 \,s$ & $0.101600 \,s$ \\ 
    $2048$ &  $2.1839\;10^{-7}$ & $1.998$ & $82.168015 \,s$ & $0.469672 \,s$
  \end{tabular}
  \caption{For uniform circular mesh, maximum relative error between numerical results and analytic solution for $10$ different values of radial cell number $N$. Poloidal mesh number are defined as $N_\theta=2 N$. CPU time is reported for the two steps of the direct Poisson solver: (i) precomputation (symbolic analysis + factorization) and (ii) solving.}
  \label{tab:mesh_uniform}
\end{table}

\subsubsection{Non-uniform mesh}
In this section, tests are now performed on non-uniform meshes defined with $N_r$ points in the radial direction and divided into $N_d$ sub-domains $\Omega_d$ depending on the number of points in the poloidal direction $N_{\theta[\Omega_d]}$, the maximum number being equal to $N_{\theta{\rm max}}=2(N_r-1)=2N$. Let $\nu_d$ denotes the reduction of poloidal points for domain $\Omega_d$ compared to the maximum number, namely $N_{\theta[\Omega_d]}=N_{\theta{\rm max}}/\nu_d$ and $\nu=(\nu_1,\cdots,\nu_{N_d})$. Each sub-domain $\Omega_d$ of $\Omega_{\rm non-uniform}$ is an annulus defined by
\begin{equation*}
  \Omega_d=\left\{(r,\theta):\left(\frac{(d-1)N}{3}+\frac{1}{2}\right)\Delta r<r<\left(\frac{d\,N}{3}+\frac{1}{2}\right)\Delta r \text{ and } 0\le \theta \le 2\pi \right\}
\end{equation*}
For all the numerical results presented in this section, the circular domain $\Omega_{\rm non-uniform}$ is divided into three sub-domains with $\nu=[4,2,1]$ (see Figure \ref{fig:non_uniform_mesh} for example with $N_r=8$). 
Two strategies have been studied to solve Poisson equation on a non-uniform mesh $\Omega_{\rm non-uniform}$. 
The first one where Poisson is solved as in the previous section \ref{sec:poisson_res_ref} on a uniform mesh $\Omega_{\rm uniform}$ of size $N\times 2 N$ where for all missing points $P_{i,j}$ (\textit{i.e} $P_{i,j}\in\Omega_{\rm uniform}$ but $P_{i,j}\notin \Omega_{\rm non-uniform}$), the RHS $R_{i,j}$ in equation \eqref{eq:poiss_polar_discrete_nonunif} are approximated by Lagrange interpolation of third order. 
The second strategy is the one for which matrix system \eqref{eq:A_matrix_nonunif}-\eqref{eq:b_nonunif} has been especially designed, \textit{i.e.} Poisson is directly solved on the non-uniform $\Omega_{\rm non-uniform}$ domain. 
The two strategies will be respectively called in the following: (i) case1: non-uniform mesh with uniform Poisson and (ii) case 2: non-uniform mesh with non-uniform Poisson. 
The number of equations is $1.6$ times bigger for case 1 than for case 2. 
Results for case 1 are summarized in Table \ref{tab:mesh_nonuniform_poiss_uniform} while results for case 2 are detailed in Table \ref{tab:mesh_nonuniform_poiss_nonuniform}, both show that the numerical scheme is still of second order. 
As expected, case 1 is simpler to implement but twice more expensive. As seen in Table \ref{tab:mesh_nonuniform_poiss_uniform} and Table \ref{tab:mesh_uniform}, CPU time for case 1 is of the same order than for the uniform mesh case because here only the CPU time related to the precomputation (symbolic analysis+factorization) and to the solving are recorded. 
So calculation of the right hand side with approximation (Lagrange interpolation for missing points) should also be taken into account in case 1.  
The maximum relative errors obtained for case 1 and case 2 are of the order of the one obtained for a uniform mesh for the smallest number of $N_\theta$ points. 

\begin{table}[!h]
  \begin{tabular}[h]{c|c|c|c|c|c|c}
    & sub-domain $\Omega_1$ & sub-domain $\Omega_2$ & sub-domain $\Omega_3$ &  \multicolumn{3}{|c}{global domain} \\  
    & $(N_\theta=2N/4)$ & $(N_\theta=2N/2)$ & $(N_\theta=2N)$ & \multicolumn{3}{|c}{$\Omega_{\rm non-uniform}$} \\
    \hline
          & \multicolumn{3}{|c|}{$\quad$} & & & \vspace{-0.2cm}\\
          & \multicolumn{3}{|c|}{}                       &       & CPU time & CPU time\\
    $N$   & \multicolumn{3}{|c|}{Maximum relative error} & Order &  precompute & solve \\
          & \multicolumn{3}{|c|}{$\quad$} & & & \vspace{-0.2cm}\\
    \hline
    & & & &\vspace{-0.2cm}\\
    $32$   &  $0.5453\;10^{-3}$ & $0.8412\;10^{-3}$ & $0.8426\;10^{-3}$ &         & $0.024073 \,s$   & $0.000555 \,s$ \\
    $128$  &  $0.3653\;10^{-4}$ & $0.5249\;10^{-4}$ & $0.5219\;10^{-4}$ & $2.002$ & $0.168538 \,s$   & $0.001374 \,s$ \\
    $512$  &  $0.2336\;10^{-5}$ & $0.3287\;10^{-5}$ & $0.3260\;10^{-5}$ & $1.998$ & $3.415027 \,s$   & $0.025067 \,s$ \\
    $2048$ &  $0.1473\;10^{-6}$ & $0.2058\;10^{-6}$ & $0.2039\;10^{-6}$ & $1.999$ & $81.058922 \,s$  & $0.452550 \,s$ \vspace{-0.2cm}\\
    & & & & & & \\
    \hline
  \end{tabular}
  \caption{For case 1 (non-uniform mesh with uniform Poisson solver): maximum relative error between numerical results and analytic solution for $4$ different values of radial cell number $N$ for the three sub-domains. The order of the numerical scheme is calculated on the global domain $\Omega_{\rm non-uniform}$ based on the maximum error on the three sub-domains. CPU time is reported for the two steps of the direct Poisson solver: (i) precomputation (symbolic analysis + factorization) and (ii) solving.}
  \label{tab:mesh_nonuniform_poiss_uniform}
\end{table}

\begin{table}[!h]
  \begin{tabular}[h]{c|c|c|c|c|c|c}
    & sub-domain $\Omega_1$ & sub-domain $\Omega_2$ & sub-domain $\Omega_3$ &  \multicolumn{3}{|c}{global domain} \\  
    & $(N_\theta=2N/4)$ & $(N_\theta=2N/2)$ & $(N_\theta=2N)$ & \multicolumn{3}{|c}{$\Omega_{\rm non-uniform}$} \\
    \hline
          & \multicolumn{3}{|c|}{$\quad$} & & & \vspace{-0.2cm}\\
          & \multicolumn{3}{|c|}{}                       &       & CPU time   & CPU time\\
    $N$   & \multicolumn{3}{|c|}{Maximum relative error} & Order & precompute & solve \\
          & \multicolumn{3}{|c|}{$\quad$} & & & \vspace{-0.2cm}\\
    \hline
    & & & &\vspace{-0.2cm}\\
    $32$   & $0.2894\;10^{-2}$ & $0.2512\;10^{-2}$ & $0.1706\;10^{-2}$ &         & $0.054278 \,s$ & $0.006446 \,s$ \\
    $128$  & $0.2132\;10^{-3}$ & $0.2021\;10^{-3}$ & $0.1344\;10^{-3}$ & $1.882$ & $0.090576 \,s$ & $0.000930 \,s$ \\
    $512$  & $0.1406\;10^{-4}$ & $0.1358\;10^{-4}$ & $0.8918\;10^{-5}$ & $1.961$ & $1.835005 \,s$ & $0.013966 \,s$ \\
    $2048$ & $0.8912\;10^{-6}$ & $0.8646\;10^{-6}$ & $0.5659\;10^{-6}$ & $1.989$ & $41.658237 \,s$ & $0.244283 \,s$ \vspace{-0.2cm}\\
    & & & & & \\
    \hline
  \end{tabular}
  \caption{For case 2 (non-uniform mesh with non-uniform Poisson solver): maximum relative error between numerical results and analytic solution for $4$ different values of radial cell number $N$ for the three sub-domains. The order of the numerical scheme is calculated on the global domain $\Omega_{\rm non-uniform}$ based on the maximum error on the three sub-domains. CPU time is reported for the two steps of the direct Poisson solver: (i) precomputation (symbolic analysis + factorization) and (ii) solving.}
  \label{tab:mesh_nonuniform_poiss_nonuniform}
\end{table}
Therefore, as a summary, we have shown in this section that Poisson equation \eqref{eq:poiss_polar} can be solved successfully on a non-uniform circular mesh with the coupling of 2D finite differences in polar coordinates and Lagrange interpolation of third order. The associated numerical scheme proposed in section \ref{sec:poisson} is of second order. Higher order for Lagrange polynomials has not been tested in this paper but the matrix system \eqref{eq:A_matrix_nonunif}-\eqref{eq:b_nonunif} could be easily generalized. In both cases, CPU time for the solve step remains quite reasonable even for the biggest meshes. This is a key point for a future implementation in \gysela{} code where the precomputation step will be only performed one times at the beginning.

\subsubsection{Mapping}
\noindent For validation of the code, we consider the following solution given in cartesian coordinates. We take here $a=1$ and $b=0.5$
$$
u(x,y) = \frac{\exp(x)+\exp(y)}{1+xy}.
$$

\begin{figure}[h!]
  \begin{minipage}[b]{0.45\linewidth}
    \hspace*{-2mm}
    \includegraphics[trim= .94cm 2cm .36cm 2cm,clip,width=1.1\linewidth]{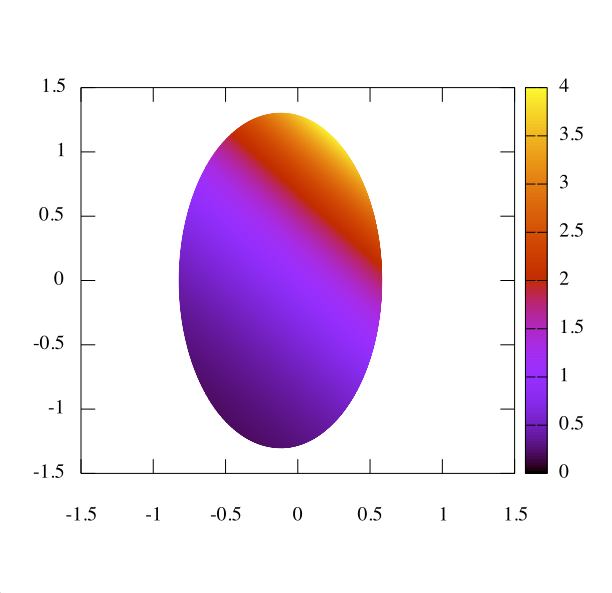}
    \caption{Analytic solution $f=2\exp(x+y)$ for the case $N=512$ and $N_\theta=1024$.\\ \ }
    \label{fig:anal_poiss_curvi_Nr512}
  \end{minipage}
  \hspace*{1cm}
  \begin{minipage}[b]{0.45\linewidth}
    \hspace*{-2mm}
    \includegraphics[trim= 1.1cm 2cm .26cm 2cm,clip,width=1.1\linewidth]{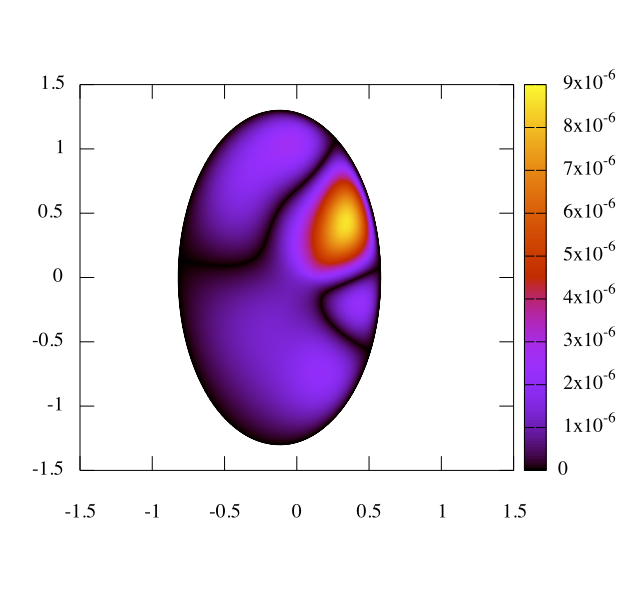}
    \caption{Error between numerical results and analytic solution for the case $N=512$ and $N_\theta=1024$ and $9$ points scheme.}
    \label{fig:error_poiss_curvi_Nr512}
  \end{minipage}
\end{figure}

We get the results in Tables \ref{table1} and  \ref{table2} using $M=2N$. It means uniform mesh along $\theta$~direction in this paragraph.
On Figures \ref{fig:anal_poiss_curvi_Nr512}, \ref{fig:error_poiss_curvi_Nr512} and Table \ref{table3}, we give numerical results for $f=2\exp(x+y)$ and large aspect ratio mapping.
We get an order two for the error as expected. We notice that the error behaves similarly for the two schemes and for the two geometries (ellipse and large aspect ratio). It seems
that the results are better for the $7$ points scheme in the case of the ellipse and for the $9$ points scheme in the case of the large aspect ratio mapping, so that there is no clear trend on what scheme would be the best in general.

\begin{table}[h!]
\begin{subtable}[b]{.45\textwidth}
\begin{tabular}[h]{l|l|l}
$N$  & Relative error  & Order   \\
    \hline
    & & \vspace{-0.2cm}\\
$4$  &   2.513D-02 (3.518D-02)    &                       \\
$8$  &     4.532D-03 (7.970D-03)    &   2.47  (2.14)               \\
$16$ &     1.352D-03 (2.041D-03)  &   1.74 (1.97)                    \\ 
$32$ & 3.346D-04 (5.185D-04) & 2.01 (1.98)\\
$64$ & 8.322D-05 (1.307D-04) & 2.01 (1.99)\\
$128$ & 2.079D-05 (3.271D-05) & 2.00 (2.00)
\end{tabular}
\caption{Ellipse}
\label{table1}
\end{subtable}
\hspace*{5mm}
\begin{subtable}[b]{.45\textwidth}
\begin{tabular}[h]{l|l|l}
%
%
$N$  & Relative error  & Order   \\
    \hline
    & & \vspace{-0.2cm}\\
$4$  &   1.038D-01 (9.249D-02)     &                       \\
$8$  &     2.941D-02 (2.385D-02)   &   1.82  (1.96)              \\
$16$ &     7.777D-03 ( 6.089D-03) &   1.92 (1.97)                   \\ 
$32$ & 2.201D-03 (1.626D-03) & 1.82 (1.90)\\
$64$ & 5.592D-04 (4.118D-04) & 1.98 (1.98) \\
$128$ &  1.407D-04 (1.037D-04) & 1.99 (1.99)
\end{tabular}
\caption{Large aspect ratio mapping}
\label{table2}
\end{subtable}
\caption{Convergence for  $7$ points \mbox{($9$~points)} scheme}
\end{table}



\begin{center}
\begin{table}
\begin{tabular}[h]{l|l|l}


$N$  & Error  & Order   \\
    \hline
    & & \vspace{-0.2cm}\\
$4$  &   0.1094    &                       \\
$8$  &     2.956D-02   &   1.89            \\
$16$ &     8.718D-03 &   1.76                  \\ 
$32$ & 2.203D-03 & 1.98 \\
$64$ & 5.526D-04 & 2.00  \\
$128$ &  1.383D-04 & 2.00 \\
$256$ &  3.460D-05 & 2.00 \\
$512$ &  8.649D-06 & 2.00 \\
$1024$ &  2.162D-06 & 2.00 \\
\end{tabular}
\caption{Convergence for large aspect ratio mapping and $9$~points scheme for $f=2\exp(x+y)$ (fortran implementation with pardiso)}
\label{table3}
\end{table}
\end{center}



\section*{Conclusion}
\label{sec:conclusion}
In the context of gyrokinetic simulation of turbulence inside a
Tokamak plasma, we have developed a strategy that incorporates an
adapted non-uniform meshing. Instead of having a circular geometry
with a uniform grid along $(r,\theta)$ dimensions to describe the
poloidal cross-section, we propose a non-uniform spacing along theta
direction aiming at computational and memory savings. Additionally, a
peculiar mapping coupled with the non-uniform mesh permits to match
more complex realistic geometry such as D-shaped plasma, thus exceeding
the former limited configuration of circular Tokamak cross-sections.
We expect this mapping to rely on analytical formulas in order to 
keep a relatively low price in term of computations, which is quite crucial
for a full-f global code as \gysela{} is.

Several features that are typically used in the gyrokinetic code
\gysela{} have been recast to handle such a new
approach. Interpolation, Advection, Gyroaverage and Poisson operators
are revisited, upgraded and analyzed in this paper. These operators have 
been studied separately. All in all, the convergence studies we provide show that
this is a workable approach that reach accuracy comparable to
uniform meshing. Adaptivity brings really a potential benefit in
term of memory savings, but it also permits to refine grid in a
specific annulus (small region along radial direction) that will be
useful for physics studies incorporating kinetic electrons. 

Future works will target the addition of this method in \gysela{}. It will
require overhauling many data structures and to combine the different operators
we have described. As this paper was not focusing on algorithms performance,
a subsequent aim will be to write parallel versions of these algorithms and to
optimize computation costs in order to compete with the execution time of the
former uniform approach. A significant gain in lowering the overall memory footprint 
is also expected.


\bibliographystyle{abbrv}
\bibliography{target.bib}

\end{document}